\documentclass{emulateapj}
\usepackage{float}
\usepackage{amsmath}
\usepackage[flushleft]{threeparttable}

\begin{document}

\title{The Coevolution of Massive Quiescent Galaxies and Their Dark Matter Halos over the Last 6 Billion Years}

\author{H. Jabran Zahid$^{1}$, Margaret J. Geller$^{1}$, Ivana Damjanov$^{2}$ \& Jubee Sohn$^{1}$ } 
\affil{$^{1}$Smithsonian Astrophysical Observatory, Center for Astrophysics Harvard \& Smithsonian - 60 Garden St., Cambridge, MA 02138, USA}
\affil{$^{2}$Department of Astronomy and Physics, Saint Mary's University - 923 Robie St., Halifax NS B2H 2C2, Canada }

\begin{abstract}
  
We investigate the growth of massive quiescent galaxies at $z<0.6$ based on the Sloan Digital Sky Survey and the Smithsonian Hectospec Lensing Survey---two magnitude limited spectroscopic surveys of high data quality and completeness. Our three parameter model links quiescent galaxies across cosmic time by self-consistently evolving stellar mass, stellar population age sensitive $D_n4000$ index, half-light radius and stellar velocity dispersion. Stellar velocity dispersion is a robust proxy of dark matter halo mass; we use it to connect galaxies and dark matter halos and thus empirically constrain their coevolution. The typical rate of stellar mass growth is $\sim \! 10 \,\, M_\odot \,\, \mathrm{yr}^{-1}$ and dark matter growth rates from our empirical model are remarkably consistent with N-body simulations. Massive quiescent galaxies grow by minor mergers with dark matter halos of mass $10^{10} \,\, M_\odot \lesssim M_{DM} \lesssim 10^{12} \,\, M_\odot$ and evolve parallel to the stellar mass-halo mass relation based on N-body simulations. Thus, the stellar mass-halo mass relation of massive galaxies apparently results primarily from dry minor merging.

\end{abstract}

\section{Introduction}

Dark matter is a mysterious substance. The large scale structure of the universe forms as density perturbations in the primordial dark matter distribution evolve under the influence of gravity. However, dark matter is not directly observable and the evolution of large scale structure and the properties of dark matter are primarily inferred from observations of galaxies and theoretical simulations. Thus, understanding how galaxies observed in the universe coevolve with their dark matter halos is fundamental. Robust observations of basic galaxy properties as a function of time combined with an observable proxy linking galaxies to their theoretical dark matter halos has great potential for shining a light on the dark matter mystery.  

Galaxy redshift surveys are an important tool of modern cosmology. Pioneering surveys of the 1980s were critical for measuring the large scale galaxy distribution and establishing the current cosmological model \citep{Davis1982, Geller1989}. Subsequently, significant observational efforts have focused on systematically exploring the galaxy distribution out to intermediate redshifts \citep[e.g.,][]{Rowan-Robinson1990, Lilly1995, Shectman1996, Cowie1996, York2000, Davis2003, Colless2003, Lilly2007, LeFevre2005, Kochanek2012, Baldry2012, Geller2014, Guzzo2014, Damjanov2018}. 

Spectroscopic data from redshift surveys yield a host of galaxy properties. Survey completeness and high data quality are important for robustly characterizing the galaxy population. Here we analyze the Sloan Digital Sky Survey (SDSS) Main Galaxy Sample \citep{Strauss2002} and the F2 field of the Smithsonian Hectospec Lensing Survey \citep[SHELS;][]{Geller2014}. Both surveys are magnitude limited and $>90\%$ complete. The data quality is sufficient to measure key galaxy properties. We supplement these spectroscopic data with SDSS and Subaru Hyper Suprime-Cam imaging \citep{Stoughton2002, Miyazaki2012}.

We analyze properties of galaxies that are directly measured or straightforwardly derived from observables to connect quiescent galaxies which have ceased star formation at $0.2<z<0.6$ to their descendants at $z<0.2$. We use stellar mass and the stellar population age sensitive $D_n4000$ index to connect the quiescent galaxy population observed in SHELS and SDSS. We constrain evolution of the population by comparing sizes and stellar velocity dispersions of galaxies in the two samples. The inclusion of stellar velocity dispersion is fundamental because it provides a robust observable link to simulations \citep{Zahid2018}.

SDSS is a foundation for studying sizes and stellar velocity dispersions of nearby galaxies \citep[e.g.,][]{Sheth2003, Shen2003, Bernardi2003c, Trujillo2004a, Trujillo2004b, Hyde2009b, Taylor2010, Bernardi2011, Sohn2017b, Zahid2017a}. These properties depend on stellar mass and stellar population age \citep{Shankar2009, Zahid2017a}. At a fixed stellar mass, quiescent galaxies in the local universe with older stellar populations are smaller and have larger stellar velocity dispersions. Similar trends are also observed at earlier times \citep{Wu2018, Damjanov2019}. The relation between stellar mass, size and stellar velocity dispersion for local quiescent galaxies is consistent with virial equilibrium expectations if the non-homologous galaxy structure is taken into account \citep{Zahid2017a}. 

Observations reveal that galaxies are significantly smaller at early times \citep{Daddi2005, Zirm2007, Buitrago2008, VanDokkum2008b, Damjanov2011, vanderwel2014, Belli2017, Damjanov2019}. In contrast, the average stellar velocity dispersion at a fixed stellar mass does not evolve significantly at $z<0.7$ \citep{Shu2012, Zahid2016c, Montero-Dorta2016}; at $z>1$ velocity dispersions appear to be larger \citep{Belli2014, Belli2017}. The lack of evolution in stellar velocity dispersion at $z<0.7$ is perhaps unexpected because observations of local galaxies imply that older quiescent galaxies observed at early times should have larger stellar velocity dispersions. 

The observed properties of quiescent galaxies appear to be inconsistent with galaxies at higher redshift passively evolving into the local population \citep[e.g.,][]{vanderwel2009, Zahid2017a}. Minor mergers may explain the measured properties of quiescent galaxies over the redshift range we examine \citep{White2007, Naab2009, Bernardi2009, Newman2012, Hilz2013}. A successful model of quiescent galaxy growth must consistently account for direct measurements of galaxy properties at earlier cosmic epochs and the dependence of these same properties on stellar population age at later times. Here we formulate such a model. 

The physical processes governing the evolution of galaxies likely reflect more fundamental processes regulated by dark matter halos which are not directly observable; their hierarchical growth and evolution are studied using N-body simulations \citep[for reviews see][]{Bertschinger1998, Dolag2008}. For a given set of initial conditions and cosmological parameters, the mass accretion histories and merger rates of individual halos can be computed directly from these simulations \citep[e.g.,][]{Mcbride2009, Fakhouri2010, vandenbosch2014}. N-body simulations have successfully reproduced our burgeoning understanding of the large scale structure traced out by galaxies \citep[e.g.,][]{Press1974, Peebles1982, Blumenthal1984, Davis1985, Springel2005}. However, connecting simulated dark matter halos to individual observed galaxies is non-trivial. 

Various approaches link galaxies to their dark matter halos \citep[for review see][]{Wechsler2018}. A standard technique matches a galaxy survey to a simulated volume of equal size and populates dark matter halos with galaxies assuming a one-to-one rank order correspondence between an observed galaxy property (e.g., luminosity, stellar mass) and dark matter halo mass \citep[i.e., abundance matching;][]{Yang2003, Kravtsov2004, Tasitsiomi2004, Conroy2006, Berrier2006, Shankar2006, Guo2010, Behroozi2013a, Kravtsov2018}. In practice, one can thus match the measured stellar mass function to the halo mass function calculated from simulations. This approach yields a relation between stellar mass and halo mass by construction. Physical interpretation of the relation often invokes uncertain star formation and baryonic feedback physics and remains ambiguous \citep{Silk2012, Wechsler2018}. 

The relation between stellar mass and halo mass is limited in constraining the coevolution of galaxies and dark matter halos partially because stellar mass is an indirect proxy. The stellar velocity dispersion of a galaxy is set by the gravitational potential; it is a directly observable property connecting galaxies to their dark matter halos \citep{Wake2012b, Schechter2015, Zahid2016c, Zahid2018}. This connection is particularly valuable because our model quantifies the change in stellar velocity dispersion of individual galaxies. We thus constrain the coevolution of galaxies and dark matter halos more directly than previous studies relying on stellar mass as the link. We interpret these previous studies using stellar velocity dispersion as an independent proxy.

We investigate the coevolution of massive quiescent galaxies and their dark matter halos over the last $\sim \! 6$ billions years of cosmic history. We describe the data in Section 2 and summarize observational constraints in Section 3. We formulate our model in Section 4, describe the fitting procedure in Section 5 and address some assumptions of our model in Section 6. In Section 7 we interpret our model results and show that they are consistent with minor merger driven growth. We use model constraints on the evolution of stellar velocity dispersion to investigate the coevolution of galaxies and dark matter halos in Section 8 and conclude in Section 9. Frequently used symbols are in Table \ref{tab:key}. We adopt the standard cosmology $(H_{0}, \Omega_{m}, \Omega_{\Lambda}) = (70$ km s$^{-1}$ Mpc$^{-1}$, 0.3, 0.7) throughout.

\section{Survey Data, Methods and Selection}

\subsection{SDSS and SHELS Data}

We analyze the Sloan Digital Sky Survey DR12\footnote{http://www.sdss.org/dr12/} \citep{Alam2015} Main Galaxy Sample of $\sim900,000$ galaxies with $r<17.8$ observed over $\sim10,000$ deg$^{2}$ in the redshift range $0\lesssim z \lesssim 0.3$ \citep{York2000}. The sample is $>90\%$ complete \citep{Strauss2002, Lazo2018}. The nominal spectral range and resolution of the SDSS observations are $3800 - 9200 \mathrm{\AA}$ and $R\sim1500$ at $5000\mathrm{\AA}$ \citep{Smee2013}, respectively. We derive stellar masses using the $ugriz$ model magnitudes measured from the SDSS imaging data \citep{Stoughton2002, Doi2010}. 

The higher redshift sample is from the Smithsonian Hectospec Lensing Survey \citep[SHELS;][]{Geller2005, Geller2014, Geller2016}. The survey covers two 4 deg$^{2}$ fields (F1 and F2) of the Deep Lensing Survey \citep[DLS;][]{Wittman2002}. Here we analyze the F2 field and refer to this throughout as the SHELS sample. 

The SHELS survey consists of $\sim13,300$ galaxies in the redshift range of $0<z<0.7$ and is $\gtrsim90\%$ complete at $R<20.6$ ($r\lesssim20.9$); a level of spectroscopic completeness comparable to the SDSS Main Galaxy Sample at $r<17.8$ \citep{Strauss2002}. The spectra are obtained with Hectospec, a 300 fiber optical spectrograph on the 6.5m MMT \citep{Fabricant2005}. The nominal spectral range and resolution of the observations are $3700-9100 \mathrm{\AA}$ and $R\sim1000$ at $5000\mathrm{\AA}$, respectively. We derive stellar masses using the SDSS $ugriz$ model magnitudes from the imaging pipeline \citep{Stoughton2002}. Redshifts, stellar masses and $D_n4000$ indices for galaxies in F2 are provided in \citet{Geller2014}. Stellar velocity dispersions and sizes of SHELS galaxies are investigated in \citet{Zahid2016c} and \citet{Damjanov2019}, respectively.

\subsection{Stellar Mass}

To derive stellar mass we estimate the mass-to-light ratio ($MLR$) by $\chi^2$ fitting synthetic spectral energy distributions (SEDs) to the observed SDSS photometry. Stellar masses derived using this approach have absolute uncertainties of $\sim0.3$ dex \citep{Conroy2009a} due to uncertainties in the star formation history (SFH), metallicity, dust extinction, stellar templates and IMF adopted to fit the SED. Our analysis relies only on the relative accuracy of stellar mass estimates. We mitigate systematic offsets by consistently calculating stellar masses for SDSS and SHELS from $ugriz$ SDSS model magnitudes using the {\sc Lephare}\footnote{http://www.cfht.hawaii.edu/$\sim$arnouts/LEPHARE/lephare.html} fitting code \citep{Arnouts1999, Ilbert2006b}. 

We fit observed SEDs using the stellar population synthesis (SPS) models of \citet{Bruzual2003} and the \citet{Chabrier2003} initial mass function (IMF). The models have an exponentially declining SFHs (star formation rate $\propto e^{-t/\tau}$) with e-folding times of $\tau = 0.1,0.3,1,2,3,5,10,15$ and $30$ Gyr and three metallicities. We adopt the \citet{Calzetti2000} extinction law and allow $E(B-V)$ to range from 0 to 0.6. The stellar population ages range between 0.01 and 13 Gyr. We generate synthetic SEDs from SPS models by varying extinction and stellar population age. Each synthesized SED is normalized to solar luminosity and stellar mass is the scale factor between the observed and synthetic SED in units of solar masses. The procedure yields a distribution for the best-fit stellar mass and we adopt the median of the distribution and denote this median as $M_\ast$.  

We compare two independent SED fitting methods and estimate a $\sim0.1$ dex dispersion in stellar mass estimates which is consistent with the observational errors \citep[e.g.,][]{Zahid2014b}. The direct proportionality between stellar masses and dynamical masses demonstrate that stellar mass estimates are robust and that the central regions of local quiescent galaxies likely have a negligible contribution from dark matter \citep{Zahid2017a}.

\subsection{Galaxy Radius}

Sizes of SDSS galaxies are measured by the NYU group \citep{Blanton2005a, Blanton2005b, Padmanabhan2008}. They fit SDSS photometry with a \citet{Sersic1968} model. The model fit yields an index $n$ describing the shape of the profile and the half-light radius. We refer to measured half-light radius throughout this work as $R$. \citet{Blanton2005b} fit the S\'ersic model to the 1D radial profile measured by taking the mean flux in annuli centered on the galaxy profile peak in all five SDSS photometric bands. To account for the seeing, the S\'ersic model is convolved with a Gaussian seeing model prior to fitting. The typical seeing is $\gtrsim1''.2$ \citep{Stoughton2002}, thus measurements of the radius are limited to about half this value. For the SDSS sample analyzed in this study, the median half-light radius is $2''.7$ and $>99\%$ have sizes larger than the typical seeing limit. For consistency with SHELS, we adopt $R$ measured in the $i$-band.

Half-light radii of SHELS galaxies and details of the procedure are published in \citet{Damjanov2019}. Here we provide a brief summary. Half-light radii are measured from Subaru Hyper Suprime-Cam \citep[HSC;][]{Miyazaki2012} $i$-band imaging of the F2 field by fitting a S\'ersic model to the two-dimensional brightness distribution of each galaxy using the SExtractor software\footnote{https://www.astromatic.net/software/sextractor} developed by \citet{Bertin1996}. The procedure yields a half-light radius along the semi-major axis, $R_{maj}$, axial ratio $b/a$ and S\'ersic index. \citet{Damjanov2019} find that while the half-light radius and axial ratio are robust to fitting procedure, the S\'ersic index measured from these deep images is not stable to initial guess adopted when fitting the profile parameters. Thus, we do not use the S\'ersic index in our analysis.

The NYU group fit the 1D azimuthally averaged profile, effectively circularizing the half-light radii for elliptical galaxies. For consistency, we circularize the half-light radii we measure for SHELS galaxies by taking $R = R_{maj} \times \sqrt{b/a}$.

{Sizes of SDSS and SHELS galaxies must be consistently measured to avoid spuriously interpreting systematic errors as redshift evolution. \citet{Damjanov2019} compare half-light radii measured for 796 SHELS galaxies which have both NYU group and circularized HSC based measurements (see their Figure 1). Using this sample, the median logarithmic difference in the two size measurements is $0.004 \pm 0.003$ dex (error on median is bootstrapped). The differences in the two size measurements are nearly normally distributed with a standard deviation $\sigma_{size} = 0.06$ dex. Assuming the observational errors in the two measurements are similar, we estimate the typical uncertainty in the size is $\sim10\%$ ($\sigma_{size}/\sqrt2 = 0.04$ dex) and is attributable to observational error. There are no statistically significant differences or offsets between the two measurements. }

\begin{table}
\begin{center}
\caption{Key for Frequently Used Symbols}
\begin{tabular}{c l}
\hline
\hline
Symbol & Definition \\

\hline
$z$ & measured redshift \\
$M_\ast$ & measured stellar mass \\
$R$ & measured half-light radius \\
$\sigma$ & measured stellar velocity dispersion \\
$D_n4000$ & measured $D_n4000$ index \\
$M_{DM}$ & dark matter halo mass of SHELS galaxies \\
                  & calculated using $\sigma$ \\
$\Delta t$ & cosmic time elapsed between two redshifts in \\
                 & years \\
$M_\ast^{e}$ & stellar mass of SHELS galaxies evolved  \\
                      &forward in time with model\\
$R^{e}$ &  half-light radius of SHELS galaxies evolved  \\
               &  forward in time with our model\\
$\sigma^{e}$ & stellar velocity dispersion of SHELS galaxies  \\
                       &evolved forward in time with our model\\
$D_n4000^{e}$ & $D_n4000$ of SHELS galaxies evolved forward in \\
                          &  time with our model \\
$M_{DM}^{e}$ & dark matter halo mass of SHELS galaxies  \\
                        & calculated using $\sigma^{e}$ \\
$M_\ast^{p}$ & stellar mass of SDSS galaxies projected back   \\
			&in time with our model\\
$R^{p}$ &  half-light radius of SDSS galaxies projected   \\
             & back in time with our model\\
$\sigma^{p}$ & stellar velocity dispersion of SDSS galaxies  \\
                     &projected back in time with our model\\
$\dot{M}_\ast$ & average stellar mass growth rate of SHELS  \\
                        & galaxies \\
$\delta M_\ast$ & total change in stellar mass of SHELS galaxies \\
$\dot{M}_{DM}$ & average dark matter growth rate of SHELS \\
                         & galaxies \\ \\

\hline
\label{tab:key}
\end{tabular}
\end{center}
\end{table}

\begin{figure*}
\begin{center}
\includegraphics[width =  2\columnwidth]{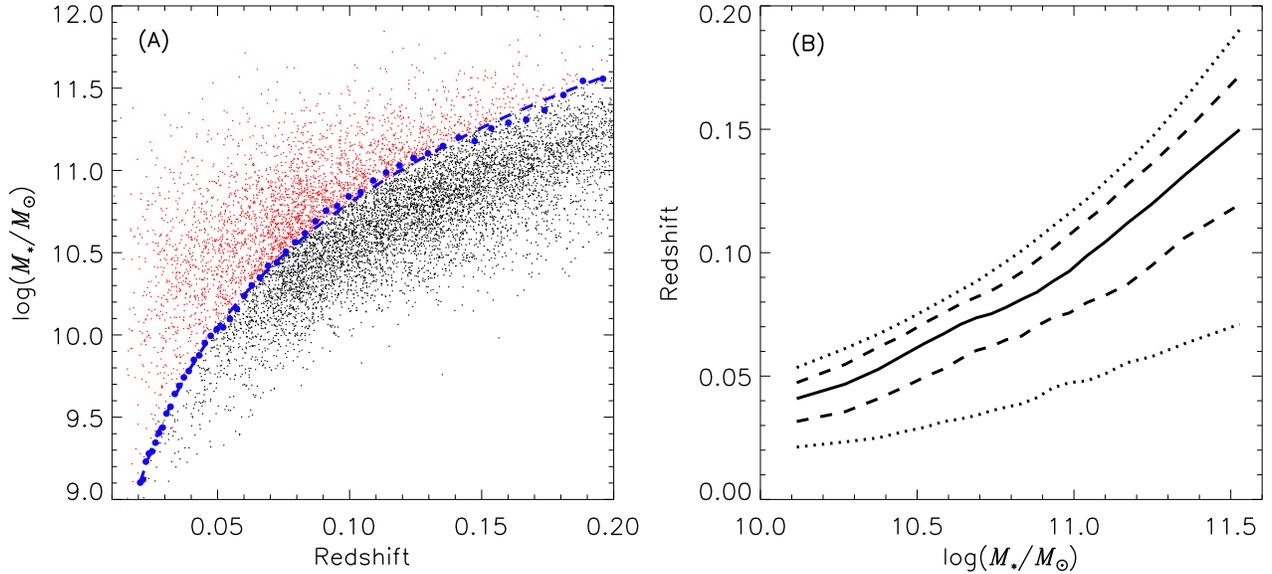}
\end{center}
\caption{ {(A) $M_\ast$ as a function of redshift for the parent SDSS sample of quiescent galaxies. Blue points and the dashed curve show the empirically derived stellar mass limit as a function of redshift. Red points are the selected sample. For clarity, only a random subset of the data are shown. (B) Solid line shows the median redshift in bins of stellar mass for the selected SDSS sample. The dashed and dotted black curves indicate limits of the central 50 and 90\% of SDSS galaxies, respectively. }}
\label{fig:sdss_selection}
\end{figure*}

\subsection{Stellar Velocity Dispersion}

The 1-dimensional line-of-sight (LOS) central stellar velocity dispersion is measured from stellar absorption lines observed through circular fiber apertures centered on each galaxy and is quoted in km s$^{-1}$. We refer to the measured LOS central stellar velocity dispersion as the stellar velocity dispersion and denote it with the symbol $\sigma$.

Stellar velocity dispersions for SDSS galaxies are measured from spectra observed through 3" fiber apertures. \citet{Thomas2013} use the \citet{Maraston2011} stellar population templates based on the Medium-resolution Isaac Newton Telescope Library of Empirical Spectra \citep[MILES;][]{Sanchez-Blazquez2006} and the Penalized Pixel-Fitting (p{\sc PXF}) code \citep{Cappellari2004}. The templates are matched to the SDSS resolution and are convolved with a range of stellar velocity dispersions and fit to the data. The best-fit stellar velocity dispersion is determined by minimizing the $\chi^2$ in the rest-frame wavelength range of $4500-6500\mathrm{\AA}$.

SHELS spectra are observed through the 1''\!\!.5 fiber aperture of Hectospec \citep[for details see][]{Fabricant2013}. Stellar velocity dispersions are measured using the University of Lyon Spectroscopic analysis Software \citep[ULySS;][]{Koleva2009}. Stellar population templates are calculated with the MILES stellar library and {\sc PEGASE-HR} code \citep{LeBorgne2004}. They are matched to the Hectospec resolution. The models are parameterized by age and metallicity and convolved with varying stellar velocity dispersions. The best-fit age, metallicity and stellar velocity dispersion are determined by minimizing the $\chi^2$ of the convolved templates and observed spectrum in the rest-frame spectral range of $4100-5500\mathrm{\AA}$ \citep{Fabricant2013}; this spectral range minimizes the error and provides the most stable measurements for Hectospec data. 

\citet{Zahid2016c} cross-calibrate the SDSS and Hectospec measurements by comparing stellar velocity dispersions measured for the same object. They find that the largest difference in the two measurements is due to the aperture and derive an empirical correction: 
\begin{equation}
\frac{\sigma_{\mathrm{SDSS}}}{\sigma_{\mathrm{HECTO}}} = \left( \frac{R_{\mathrm{SDSS}}}{R_{\mathrm{HECTO}}} \right)^\beta, 
\label{eq:apcorr}
\end{equation}
where $R_{\mathrm{SDSS}} = 1''\!\!.5$ and $R_{\mathrm{HECTO}} = 0''\!\!.75$ are the fiber aperture {radii}. They find $\beta = -0.033 \pm  0.011$ minimizes the difference between the SDSS and SHELS measurements. This value of $\beta$ is consistent with similar corrections derived in the literature \citep{Jorgensen1995, Mehlert2003, Cappellari2006}. 


To consistently compare SDSS and SHELS we correct stellar velocity dispersion to the half-light radius using Equation \ref{eq:apcorr}. The aperture corrections applied to both samples are small, ranging between $\pm0.02$ dex.

\begin{figure*}
\begin{center}
\includegraphics[width = 2 \columnwidth]{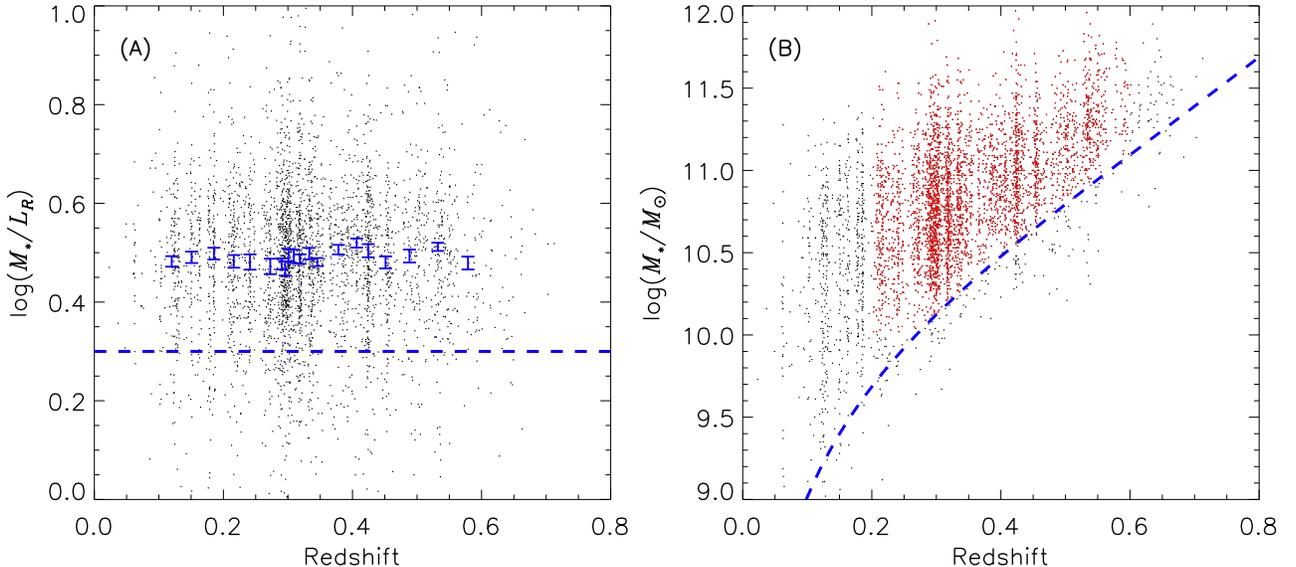}
\end{center}
\caption{ {(A) $R$-band $MLR$ as a function of redshift for the parent SHELS sample of quiescent galaxies. Error bars show the median $MLR$ in 20 equally populated redshift bins. Errors are bootstrapped. The $MLR$ is independent of redshift. The blue dashed line is the fiducial $MLR$ we adopt to translate the SHELS magnitude limit to a stellar mass limit. (B) $M_\ast$ as a function of redshift for the parent SHELS sample of quiescent galaxies. The blue dashed curve is the stellar mass limit assuming the constant $MLR$ in (A). Red points indicate the selected sample.} }
\label{fig:shels_selection}
\end{figure*}

\subsection{$D_{n}4000$ Index}

The $D_n4000$ index is a spectroscopic indicator measured from an optical spectrum quantifying the 4000$\mathrm{\AA}$ break \citep{Balogh1999}. Specifically, it is the average flux density (in frequency units) measured between 4000-4100$\mathrm{\AA}$ relative to the average flux measured between 3850-3950$\mathrm{\AA}$. For SDSS galaxies, we adopt measurements from the MPA/JHU\footnote{http://wwwmpa.mpa-garching.mpg.de/SDSS/DR7/} \citep{Kauffmann2003a}. We take $D_n4000$ indices for SHELS galaxies from \citet{Geller2014}. \citet{Fabricant2008} compare $D_n4000$ indices measured with SDSS and Hectospec for an overlapping sample of galaxies. The two measurements are consistent to within a few percent which is sufficient for our analysis. Throughout the remainder of the paper we refer to the measured $D_n4000$ index simply as $D_n4000$.

{$D_n4000$ is sensitive to stellar population age and increases monotonically after star formation ceases. The index distribution is bimodal for galaxies in the local and intermediate redshift universe because quiescent galaxies are dominated by older stars \citep[e.g.,][]{Kauffmann2003a, Geller2014}. \citet{Damjanov2018} demonstrate that $D_n4000$ can be used to identify quiescent galaxies robustly (see their Figure 8). }

{We use $D_n4000$ to select quiescent galaxies where stellar kinematics are typically dominated by random motions \citep[e.g.,][]{Brinchmann2004}. We also use $D_n4000$ as an evolutionary link between galaxies in the local universe and their younger progenitors at higher redshift. }

We limit the maximum $D_n4000 = 2.05$. The half-light radii and velocity dispersions no longer depend on $D_n4000$ for $D_n4000 \gtrsim2.05$; the dependence on $D_n4000$ saturates. $D_n4000$ can be systematically biased due to poor sky subtraction and large values likely reflect systematic errors affecting a small fraction of galaxies. We mitigate the impact of $D_n4000$ saturation and outliers by limiting $D_n4000$ to 2.05; we set values of $D_n4000>2.05$ to this limiting value.


\subsection{Sample Selection}

We require $D_n4000 > 1.5$ to select galaxies dominated by older stellar populations which have ceased star formation and have stellar kinematics dominated by random stellar motions. Kinematics dominated by random motion are critical for interpreting the stellar velocity dispersion as a virial quantity \citep[e.g.,][]{Zahid2017a, Zahid2018}. The $D_n4000 = 1.5$ limit is identified based on the bimodality of the distribution which separates star-forming and quiescent galaxies \citep[e.g.,][]{Kauffmann2003a, Woods2010, Geller2014, Geller2016}. Various results based on this $Dn4000$ selection criterion are presented in previous studies \citep[e.g.,][]{Utsumi2016, Zahid2016c, Sohn2017a, Sohn2017b, Zahid2017a, Sohn2018, Damjanov2018, Damjanov2019}.

SDSS is our reference sample. We take stellar mass as the independent variable and select a stellar mass {limited} sample in Section 2.6.1 to derive unbiased relations for a representative sample of galaxies in the local universe. Failure to derive unbiased relations could yield spurious conclusions regarding the evolution of quiescent galaxies. 

{The SHELS sample selection described in Section 2.6.2 differs from the one for SDSS. The SDSS selection yields an unbiased set of relations from a sample spanning a large relative redshift range. The lower redshift limit of the SDSS sample is a factor of 20 smaller than the upper limit. In contrast, the lower redshift limit of the SHELS sample is only a factor of 3 smaller than the upper limit.}

{We use SHELS galaxies as individual test particles. We derive SHELS relations in Section 3 for illustrative purposes only. We model the evolution of individual SHELS galaxies and determine where each SHELS galaxy lands on the fiducial SDSS relations. We constrain our evolutionary model by minimizing the difference between the evolved SHELS galaxies and the fiducial SDSS relations.}

\subsubsection{Sloan Digital Sky Survey}

We  examine SDSS galaxies over a relatively large redshift range of $0.01<z<0.2$ because massive galaxies are rare. We limit the analysis to galaxies with $M_\ast > 10^{10} M_\odot$.

We take stellar mass as the independent variable and construct a sample that is ``stellar mass {limited}." At any given redshift, all galaxies above the stellar mass limit are included out to that redshift regardless of their $MLR$. A similar sample selection is applied in the derivation of stellar mass \citep[e.g.,][]{Fontana2006, Perez-Gonzalez2008, Marchesini2009, Weigel2016} and stellar velocity dispersion \citep{Sohn2017b} functions.

\begin{figure*}
\begin{center}
\includegraphics[width = 2 \columnwidth]{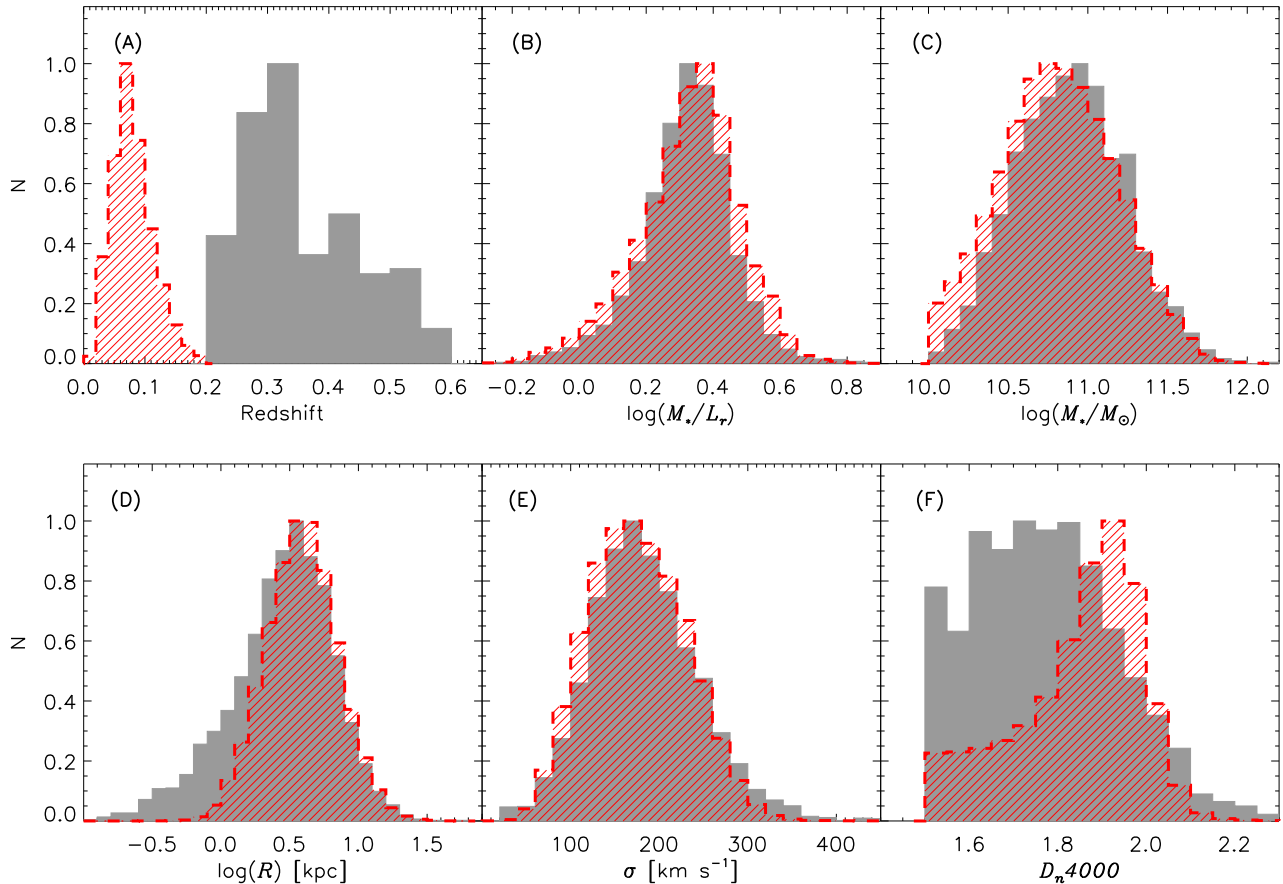}
\end{center}
\caption{ {(A) Redshift, (B) $r$-band $MLR$, (C) $M_\ast$, (D) $R$, (E) $\sigma$ and (F) $D_n4000$ distributions of the selected SDSS (red hatched) and SHELS (gray) samples. The SDSS and SHELS are $r$-band and $R$-band selected, respectively. For consistency we show the $r$-band $MLR$ distribution in (B) for both the SDSS and SHELS sample.} }
\label{fig:hist}
\end{figure*}

The SDSS survey is magnitude limited in the $r$-band. Thus, the absolute magnitude limit of galaxies in the survey are a function of redshift; a fixed limit in apparent brightness yields a sample in which only intrinsically brighter galaxies are observed at the higher redshifts probed by the survey. The standard approach of deriving a volume or distance limited sample yields a sample which is complete to a fixed absolute magnitude. In other words, galaxies brighter than the absolute magnitude limit are seen throughout the volume. However, such a sample is not \emph{volume} limited in other galaxies properties due to the scatter between various galaxy properties and the $r$-band absolute magnitude. To derive a stellar mass {limited} sample, we must account for the scatter between $r$-band absolute magnitude and stellar mass.

We empirically derive the stellar mass limit for the SDSS sample in direct analogy to the procedure described in \citet{Sohn2017b} for deriving the stellar velocity dispersion limit. We derive the limit at a given redshift as follows: 1) we derive a volume limited sample from the K-corrected absolute $r$-band magnitude using the standard approach; 2) for this volume limited sample, the upper and lower stellar mass limits containing 95\% of galaxies are empirically determined as a function of absolute $r$-band magnitude; 3) a linear model is fit to the upper envelope of the stellar mass distribution as a function of absolute magnitude; 4) the upper stellar mass limit at the limiting absolute magnitude is derived from the fit. We repeat this procedure at each redshift to calculate the stellar mass limit as a function of redshift and show the results in Figure \ref{fig:sdss_selection}A.

SDSS is a magnitude limited survey and thus our stellar mass selection criteria is explicitly a function of redshift. {Figure \ref{fig:sdss_selection}B shows the median redshift of the selected sample as a function of stellar mass.} For galaxies with $M_\ast = 10^{10.5} M_\odot$ and $M_\ast = 10^{11.5} M_\odot$, the median redshift is 0.05 and 0.15, respectively. We account for the sample redshift dependence in our analysis.

The final stellar mass {limited} SDSS sample is comprised of $\sim110,000$ galaxies. Nearly all galaxies in the SDSS have data quality sufficient for measuring $D_n4000$, half-light radius and stellar velocity dispersion. Figure \ref{fig:hist} shows the properties of the final SDSS sample.

\subsubsection{Smithsonian Hectospec Lensing Survey}

\begin{figure*}
\begin{center}
\includegraphics[width = 1.77 \columnwidth]{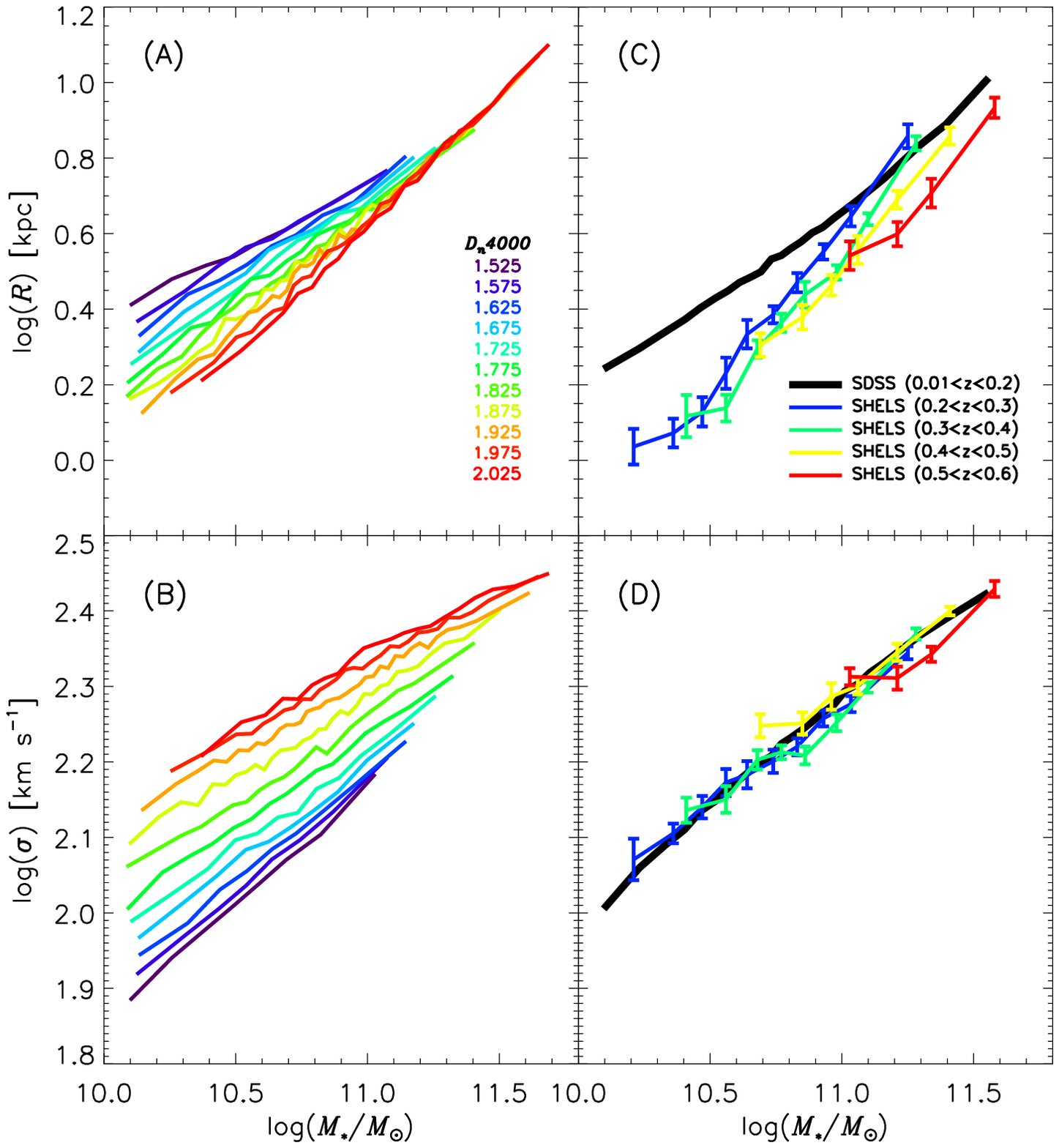}
\end{center}
\caption{(A) Median $R$ in bins of $M_\ast$ and $D_n4000$ for the SDSS sample. The median is calculated by first sorting data into $D_n4000$ bins spaced by $\Delta D_n4000 = 0.05$ and then into evenly populated $M_\ast$ bins such that all bins contain $>500$ galaxies. At a fixed $M_\ast$, $D_n4000$ is anti-correlated with $R$. (B) Median $\sigma$ in bins of $M_\ast$ and $D_n4000$ for the SDSS sample. The binning procedure is the same as in (A). At a fixed $M_\ast$, $D_n4000$ is correlated with $\sigma$. (C) Median $R$ in bins of $M_\ast$ for SDSS (thick black curve) and SHELS (colored curves and error bars) galaxies. There is clear evolution in the relation. {However, the apparent differential redshift evolution of $R$ with respect to $M_\ast$ between SHELS and SDSS is enhanced due to the large redshift range of the SDSS sample which varies with stellar mass as seen in Figure \ref{fig:sdss_selection}B (see Sections 2.6.1 and 3 for more details).} (D) Median $\sigma$ in bins of $M_\ast$ for SDSS (thick black curve) and SHELS (colored curves and error bars) galaxies. The relation does not appear to evolve significantly. Errors in (C) and (D) are bootstrapped.}
\label{fig:constraints}
\end{figure*}

We analyze quiescent galaxies which are part of the ``complete" magnitude limited SHELS sample over the redshift range $0.2<z<0.6$. The lower redshift limit prevents overlap with the SDSS sample. The upper redshift limit is applied because for the given magnitude limit of the survey the sample becomes sparse at $z>0.6$ \citep[see Figure 4 of ][]{Zahid2016c}. We limit the analysis to galaxies with $M_\ast > 10^{10} M_\odot$.

{We remove galaxies with large stellar mass errors by translating the magnitude limit into a stellar mass limit adopting a constant $MLR$ to convert from $R$-band absolute magnitude and stellar mass. This assumption is reasonable for $D_n4000 > 1.5$ galaxies \citep[see Figure 12 in][]{Geller2014}. Figure \ref{fig:shels_selection}A shows that the $MLR$ is independent of redshift for the SHELS sample and has a small dispersion (0.11 dex; see Figure \ref{fig:hist}B). Figure \ref{fig:shels_selection}B shows the stellar mass limit as a function of redshift. We remove galaxies below the stellar mass limit because of probable large errors in the stellar mass. }

{We test whether our primary results depend on the fiducial $MLR$ we adopt to select SHELS galaxies (i.e., $MLR = 0.3$) by testing a range of $MLR$s. Our primary results are insensitive to the choice of $MLR$. Uncertainties in our results are dominated by statistical errors (see Table \ref{tab:param_appendix} and the appendix for details).}

Our SHELS sample consists of 3644 galaxies. 92\% of SHELS galaxies meeting the selection criteria have measurements of stellar mass, size, stellar velocity dispersion and $D_n4000$. The SHELS parent sample is 95\% complete to the magnitude limit. Thus, we estimate that our selected sample is $\sim90\%$ complete to the magnitude limit. The properties of the final SHELS sample are shown in Figure \ref{fig:hist}.

{Figure \ref{fig:hist} reveals important observational trends which constrain our model of galaxy growth. Mass-to-light ratio variations are small for quiescent ($D_n4000 > 1.5$) galaxies. Quiescent galaxies at $0.2<z<0.6$ are smaller in size than local galaxies (Figure \ref{fig:hist}D) and have a velocity dispersion distribution that is broadly consistent with the local population (Figure \ref{fig:hist}E). The $D_n4000$ distributions in Figure \ref{fig:hist}F reflect the younger stellar populations of SHELS galaxies as compared to the local population. Our model constrains the growth of quiescent galaxies by transforming the SHELS population onto the SDSS population using these observable properties and trends in Figure \ref{fig:hist} as constraints. }

\section{Observational Constraints for Quiescent Galaxy Evolution}

The size-mass ($MS$) and stellar velocity dispersion-mass ($M\sigma$) relations observationally constrain our model of quiescent galaxy evolution. Figures \ref{fig:constraints}A and \ref{fig:constraints}B show that for galaxies in the local universe, half-light radius and stellar velocity dispersion are correlated with stellar mass and $D_n4000$. At a fixed stellar mass, galaxies with older stellar populations are smaller and have larger stellar velocity dispersions. 

Figure \ref{fig:constraints}C shows the $MS$ relation as a function redshift. At a fixed stellar mass, galaxies at earlier times are smaller as compared to the local population and the difference in size is greatest for less massive galaxies. Figure \ref{fig:constraints}A shows that the shallower slope of the SDSS $MS$ relation at low stellar masses results from young quiescent galaxies which are larger than the older quiescent galaxy population. We note that redshift range probed by the SDSS sample is a function of stellar mass; SDSS galaxies at low and high masses are typically at $z\sim0.05$ and $z\sim0.15$, respectively (see Figure \ref{fig:sdss_selection}B and Section 2.6.1). This redshift dependence on stellar mass for SDSS is an additional effect enhancing the inherent differential evolution of galaxy half-light radius with respect to stellar mass \citep[e.g.,][]{Damjanov2019}. In contrast to the $MS$ relation, Figure \ref{fig:constraints}D shows that the $M\sigma$ relation does not significantly evolve with redshift \citep[for more details see][]{Zahid2016c}.

Our primary aim is to understand how quiescent galaxies at $0.2<z<0.6$ evolve into the local population at $z<0.2$. We seek a quantitative model of quiescent galaxy evolution which links half-light radii and stellar velocity dispersions of galaxies between different epochs while simultaneously accounting for the dependence of these properties on the stellar population age as observed in local galaxies. To achieve this goal, we formulate a phenomenological model that transforms the SHELS population into the SDSS population using stellar mass and $D_n4000$ as an evolutionary link thus self-consistently satisfying the observational constraints.

\section{A Quantitative Model to Link Quiescent Galaxies Across Cosmic Time}

We develop a model to quantify quiescent galaxy evolution using observations from SHELS and SDSS. To facilitate the comparison between samples, we fit the local data in Section 4.1. We use $D_n4000$ as an evolutionary link and in Section 4.2 we quantify its evolution with a stellar population synthesis model. In Section 4.3 we show that passive evolution alone can not account for the properties of quiescent galaxies observed at different cosmic epochs. We derive our evolutionary model in Section 4.4.

\subsection{The Local Relation Between Stellar Velocity Dispersion, Size, Stellar Mass and $D_{n}4000$}

\begin{figure*}
\begin{center}
\includegraphics[width = 2 \columnwidth]{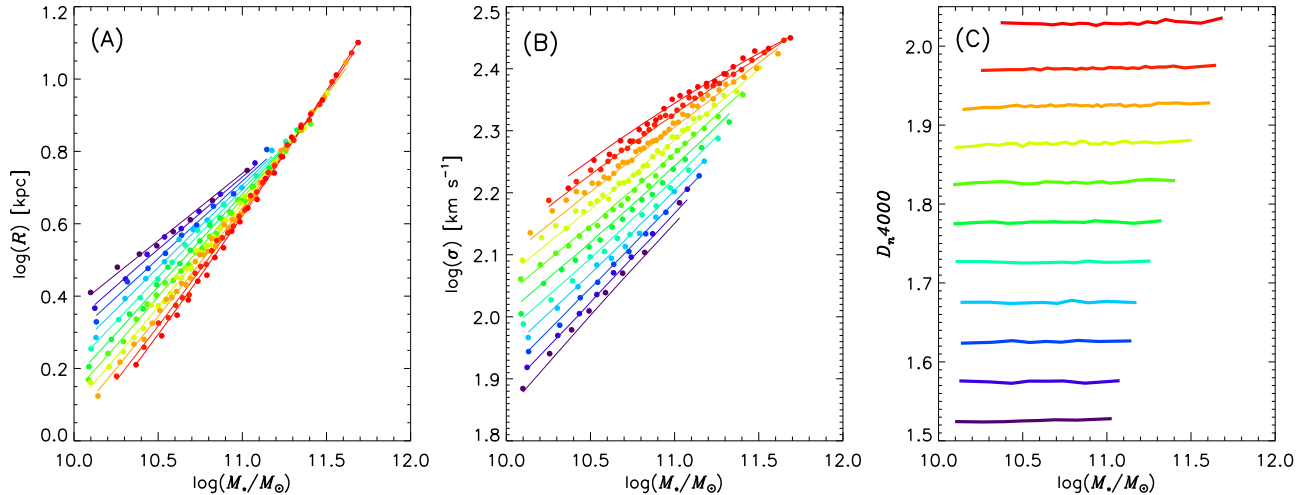}
\end{center}
\caption{(A) $R$ and (B) $\sigma$ in bins of $M_\ast$ and $D_n4000$ for the SDSS sample. These data are the same as in Figure \ref{fig:constraints}A  and \ref{fig:constraints}B. Solid curves are fits to the data (Section 4.1). (C) Median $D_n4000$ corresponding to the binned data in (A) and (B).}
\label{fig:vdisp_rad_fit}
\end{figure*}

We quantify the properties of local galaxies by binning SDSS data into 10 equally spaced bins in $D_n4000$ with $\Delta D_n4000 = 0.05$ and then into equally populated bins of stellar mass ensuring that all bins have $>500$ galaxies. We calculate the median half-light radius, stellar velocity dispersion and $D_n4000$ in each bin and plot the results in Figures \ref{fig:vdisp_rad_fit}A, \ref{fig:vdisp_rad_fit}B and \ref{fig:vdisp_rad_fit}C, respectively. The errors are bootstrapped. 

The solid lines in Figures \ref{fig:vdisp_rad_fit}A and \ref{fig:vdisp_rad_fit}B are the best-fits to the $MS$ and $M\sigma$ relations sorted by $D_n4000$, respectively. We fit the relations by minimizing $\chi^2$ using \emph{mpcurvefit.pro} implemented as part of the {\sc MPFIT IDL} package \citep{Markwardt2009}. The relations are:
\begin{equation}
\mathrm{log}(R_{SDSS}) = p_0 + p_1 \, x + p_2 \, x^2
\label{eq:sdss_radius}
\end{equation}
and
\begin{equation}
\mathrm{log}(\sigma_{SDSS}) = p_3 + p_4 \, x + p_5 \, x^2
\label{eq:sdss_sigma}
\end{equation}
where $x = \mathrm{log}(M_\ast/M_\odot)$. Each parameter $p_i$ (i = 0...5) is a function of $D_n4000$;
\begin{equation}
p_i = a_i + b_i \, y + c_i \, y^2
\end{equation}
where $y$ is the measured $D_n4000$. The fits are accurate to $\lesssim0.01$ dex which is sufficient for our application.

\subsection{Connecting Galaxies with $D_{n}4000$}

\begin{figure}
\begin{center}
\includegraphics[width =  \columnwidth]{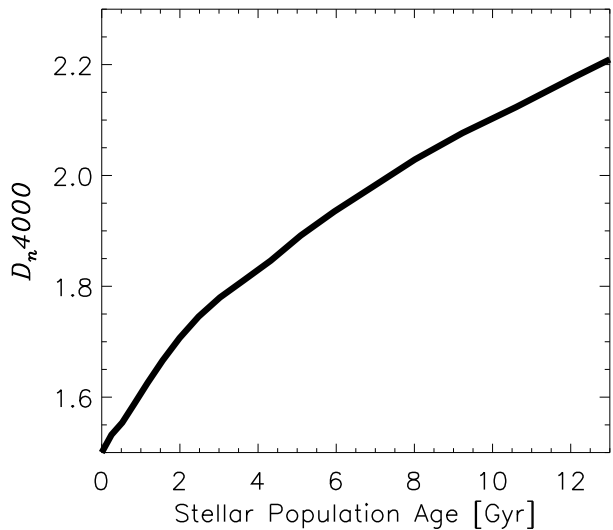}
\end{center}
\caption{Evolution of $D_n4000$ as a function of time calculated for the fiducial FSPS model with solar metallicity and constant SFR for 1 Gyr.}
\label{fig:dn4000_model}
\end{figure}

 \begin{figure*}
\begin{center}
\includegraphics[width = 2 \columnwidth]{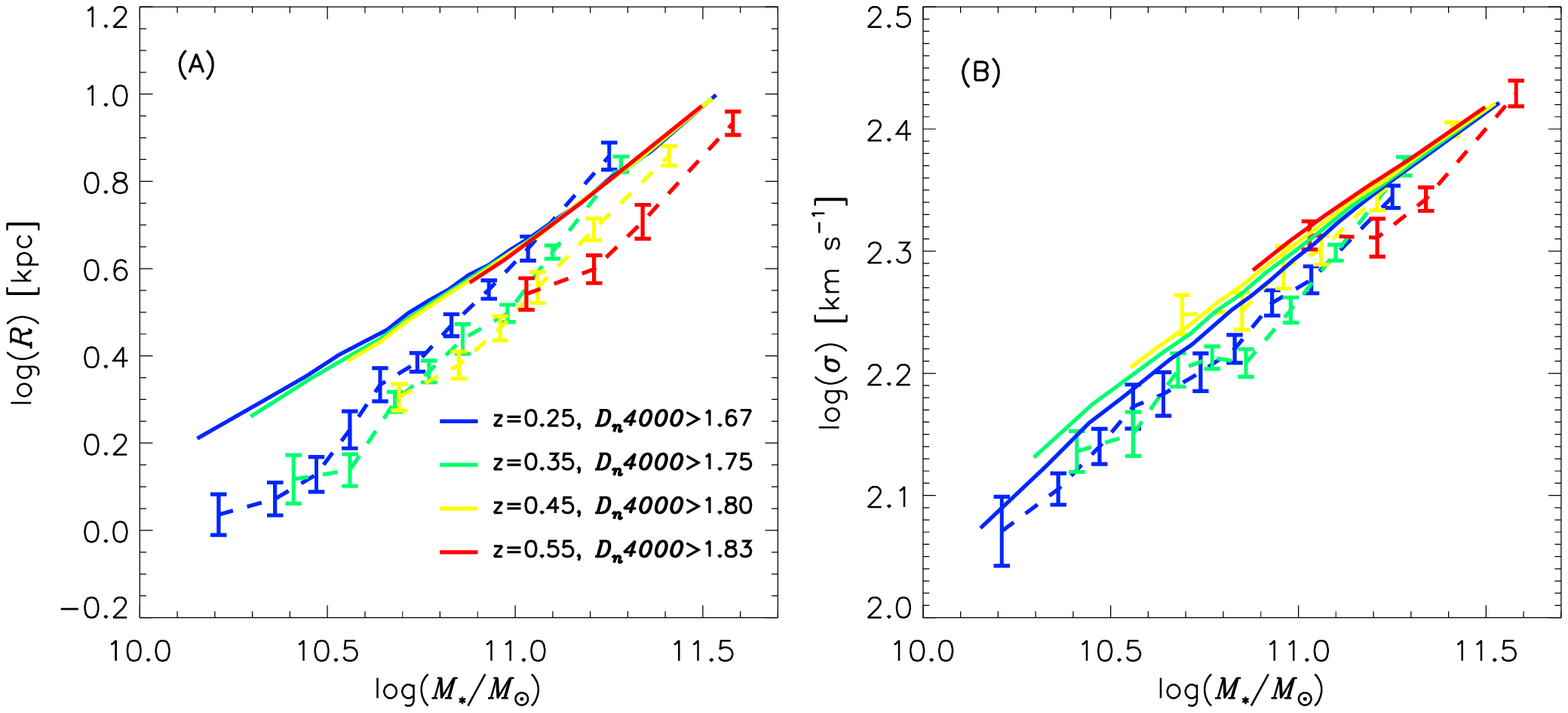}
\end{center}
\caption{Properties of SDSS galaxies assuming purely passive evolution compared with SHELS. We project quiescent galaxies back in time by selecting objects with $D_n4000$ greater than a threshold value determined from our $D_n4000$ model (Figure \ref{fig:dn4000_model}). $D_n4000$ selection thresholds corresponding to different redshifts are indicated. (A) Median $R$ and (B) $\sigma$ in bins of $M_\ast$ for SDSS projected back in time assuming purely passive evolution (solid curves) compared to direct measurements of median (A) $R$ and (B) $\sigma$ in bins of $M_\ast$ from SHELS (dashed curves and error bars). Error bars are bootstrapped. Projected relations are inconsistent with direct measurements.}
\label{fig:size_vdisp_noevol}
\end{figure*}

$D_n4000$ increases monotonically for a passively aging stellar population. We quantify this evolution with the Flexible Stellar Population Synthesis (FSPS; v3.1) model \citep{Conroy2009a, Conroy2010} implemented with a \citet{Chabrier2003} initial mass function, the MILES stellar library and the Mesa Isochrones and Stellar Tracks \citep{Dotter2016, Choi2016}.

FSPS generates spectra as a function of time given an input star formation and metallicity history. For simplicity, we adopt a constant star formation rate for 1 Gyr and solar metallicity \citep[see][]{Zahid2015}. We calculate $D_n4000$ directly from model spectra using the same procedure that we apply to observed spectra. 

Massive quiescent galaxies at $z\lesssim1$ are observed to have metallicities between solar and twice-solar \citep[e.g.,][]{Thomas2005, Choi2014, Saracco2018}. {Evolution of $D_n4000$ depends somewhat on stellar population metallicity. Thus, to test systematic dependence of our results on FSPS model metallicity, we use model spectra generated with a constant star formation rate for 1 Gyr and twice-solar metallicity.} Our results are not sensitive to the choice of inputs to FSPS; statistical errors dominate the uncertainty of our model results {(see Table \ref{tab:param_appendix} and the appendix for details)}. 


We use $D_ n4000$ as an evolutionary link between galaxies at different epochs. Figure \ref{fig:dn4000_model} shows $D_n4000$ as a function of time. We assume that once a galaxy ceases star formation and becomes quiescent, i.e. $D_n4000 > 1.5$, $D_ n4000$ evolves as shown in Figure \ref{fig:dn4000_model}. Under this assumption, $D_n4000$ is a directly measured proxy of the stellar population age and can be used to link galaxies in the local universe with their progenitors at earlier times and vice versa. For example, a galaxy that has a $D_n4000 \approx1.7$ ceased star formation ($D_n4000 = 1.5$) 2 Gyrs in the past. Conversely, a galaxy that ceases star formation at $z=0.5$ ($\sim 5$ Gyrs ago) has a $D_n4000 \approx 1.9$ at $z=0$.

\subsection{Passive Evolution Alone Can Not Explain Observations of Quiescent Galaxies}

We test the null hypothesis that quiescent galaxies cease star formation and then merely passively evolve \citep[see also][]{Zahid2017a}. By passive evolution, we mean a galaxy evolves solely because the stellar population ages. In this scenario, the quiescent population of any earlier epoch is fully represented in the galaxy population observed at some later time and stellar masses, half-light radii and stellar velocity dispersions of quiescent galaxies do not change. In this case, the observed evolution of the $MS$ relation would be due solely to the continual addition of new objects to the quiescent galaxy population with average properties differing from the pre-existing population, i.e. progenitor bias \citep{vanDokkum2001, Carollo2013}. 

$D_n4000$ encodes the epoch when star formation ceases. Assuming the null hypothesis of purely passive evolution, a larger $D_n4000$ selection threshold applied to the SDSS sample selects the quiescent galaxy population of earlier epochs. We can thus ``project" the local quiescent galaxy population back in time. For example, a galaxy that has a $D_n4000 = 1.5$ at $z=0.35$ has a $D_n4000 = 1.75$ at $z=0.1$ (median redshift of SDSS sample). If we select all quiescent galaxies in our SDSS sample with $D_n4000 > 1.75$ assuming purely passive evolution, we should recover the quiescent galaxy population at $z=0.35$.

Figure \ref{fig:size_vdisp_noevol} shows the $MS$ and $M\sigma$ relations for the projected SDSS sample compared to direct measurements from SHELS. The figure demonstrates that projection of the local quiescent galaxy population is inconsistent with direct measurements; the null hypothesis of purely passive evolution does not account for the observations.

Using a similar approach, we can evolve SHELS galaxies forward in time assuming passive evolution. The only property that changes is $D_n4000$ which evolves according to the model shown in Figure \ref{fig:dn4000_model}. We refer to the evolved $D_n4000$ with a superscript ``e", i.e. $D_n4000^e$. We evolve $D_n4000$ of SHELS galaxies to the redshift of SDSS and sort galaxies in bins of stellar mass and $D_n4000^e$. Figure \ref{fig:vdisp_rad_F2} shows the results. Compared to SDSS galaxies, half-light radii and stellar velocity dispersions of SHELS galaxies are too small. Properties of the SDSS and SHELS samples may be reconciled only if there are physical processes acting on the quiescent galaxy population in addition to purely passive evolution.

\subsection{A Model of Quiescent Galaxy Growth}

\begin{figure*}
\begin{center}
\includegraphics[width = 2 \columnwidth]{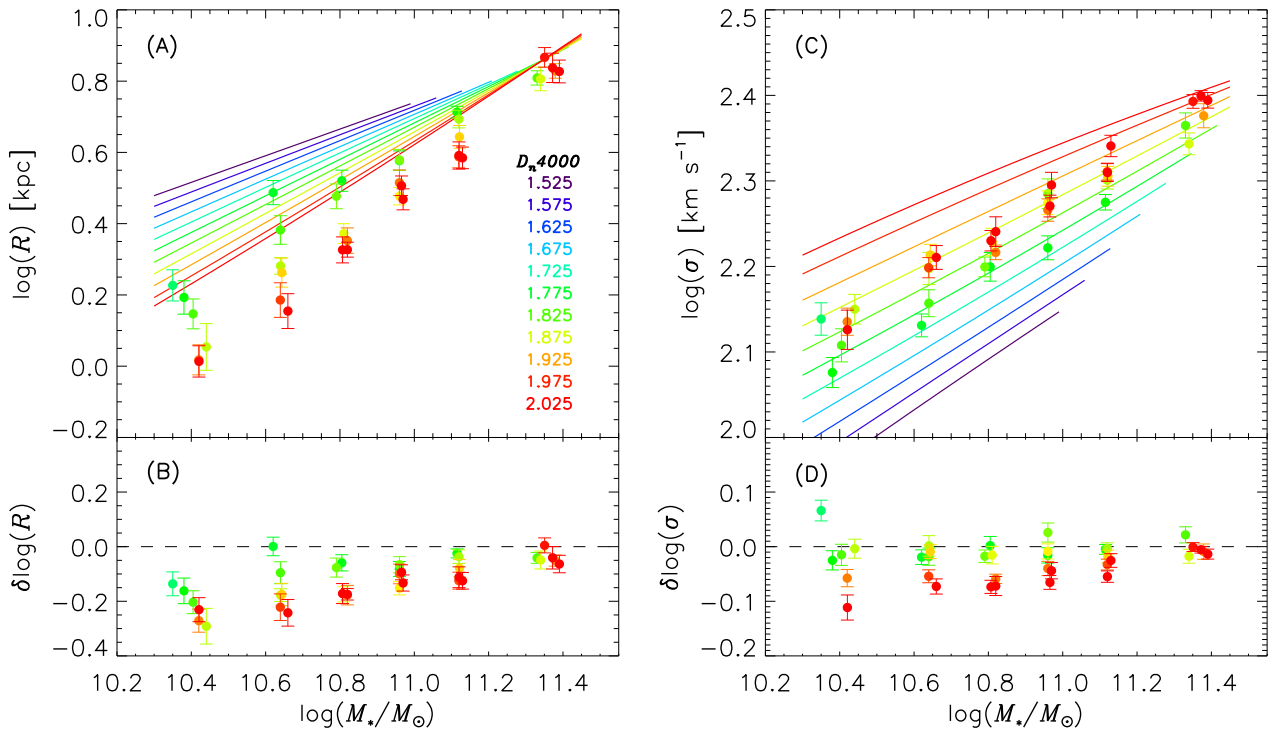}
\end{center}
\caption{Properties of SHELS galaxies assuming purely passive evolution compared with SDSS. (A) Median $R$ in bins of $M_\ast$ and $D_n4000^e$ for SHELS (points with error bars) compared to the SDSS sample (colored lines). The colored lines are the fits to the SDSS data (Equation \ref{eq:sdss_radius}). The colors of both the lines (SDSS) and the data points (SHELS) reflect the $D_n4000$ value as indicated in the legend, e.g., red lines and points correspond to galaxies with $D_n4000\sim2$. (B) Difference between $R$ from SHELS and SDSS. (C) Median $\sigma$ in bins of $M_\ast$ and evolved $D_n4000$ for SHELS (points with error bars) compared to the SDSS sample (colored lines). The colored lines are the fits to the SDSS data (Equation \ref{eq:sdss_sigma}). (D) Difference between $\sigma$ from SHELS and SDSS. $R$ and $\sigma$ of SHELS galaxies are smaller than SDSS galaxies at fixed $M_\ast$ and $D_n4000$. Error bars are bootstrapped}
\label{fig:vdisp_rad_F2}
\end{figure*}

Figure \ref{fig:vdisp_rad_F2} shows that both half-light radii and stellar velocity dispersions of SHELS galaxies must increase after they become quiescent. The only viable mechanism we are aware of for increasing radii and stellar velocity dispersions is merger/accretion driven growth \citep[e.g.,][]{Hopkins2010}. In this growth scenario, stellar mass, half-light radius and stellar velocity dispersion can all change. We construct a model for this change which we apply to the {individual galaxies in the} SHELS sample. We define the change as:
\begin{equation}
 \Delta \mathrm{log}(M_\ast)  = \mathrm{log} \left( \frac{M_\ast^e}{M_\ast} \right), 
 \end{equation}
 \begin{equation}
 \Delta \mathrm{log}(R) = \mathrm{log} \left( \frac{R^e}{R} \right),
 \end{equation}
 and
 \begin{equation}
\Delta \mathrm{log}(\sigma) = \mathrm{log} \left( \frac{\sigma^e}{\sigma} \right).
\end{equation}
Here, as with $D_n4000^e$, the superscript ``e" indicates the property evolved forward in time.  

We parameterize the evolved stellar mass as
\begin{equation}
M_\ast^e = M_\ast + \delta M_\ast \,\,\,[M_\odot]
\label{eq:mprime}
\end{equation}
where $\delta M_\ast$ is the total change in stellar mass given by
\begin{equation}
\delta M_\ast = \dot{M}_\ast \,\, \Delta t \,\,\, [M_\odot]. 
\label{eq:delta_m}
\end{equation}
Here $\Delta t$ is the amount of time that elapses between galaxies at two epochs and $\dot{M}_\ast$ is the stellar mass growth rate. The growth rate is
\begin{equation}
\dot{M}_\ast = \epsilon (D_{n}4000 - 1.5) (1+z)^3 \,\, [M_\odot \, \, \mathrm{yr}^{-1}].
\label{eq:m_acc_rate}
\end{equation}
Here $D_n4000$ and $z$ are measured from SHELS data and $\epsilon$ is a free parameter of the model which we refer to as the ``growth efficiency". 

In summary, we parameterize the change in stellar mass by the growth rate which depends on the observed $D_n4000$ and redshift of each galaxy in SHELS and a ``growth efficiency" factor $\epsilon$. We calculate the total change in stellar mass by integrating the growth rate over time. 

We parameterize the changes in half-light radius and stellar velocity dispersion with power law dependencies on the change in stellar mass:
\begin{equation}
\Delta \mathrm{log}(R) = \alpha \Delta \mathrm{log}(M_\ast)
\label{eq:rprime}
\end{equation}
and
\begin{equation}
\Delta \mathrm{log}(\sigma) = \beta \Delta \mathrm{log}(M_\ast).
\label{eq:sprime}
\end{equation}
Here $\alpha$ and $\beta$ are free parameters. 

The model we derive has three free parameters $\epsilon$, $\alpha$ and $\beta$. We forward fit the SHELS data evolved according to the model to find the best-fit. Before describing details of the fitting procedure, we discuss the derivation of our model.

We empirically derive the particular model formulation adopted in this study after implementing dozens of different parameterizations. We first fit models where the growth of stellar mass is a function of stellar mass and redshift. A striking feature of these early models is that the resulting change in stellar mass consistently implies a very narrow range of stellar growth rates. As a consequence, we parameterize the change in stellar mass by the growth rate. The growth rate appears to be a function of stellar mass and redshift. We parameterize the redshift dependence as $\dot{M}_\ast \propto (1+z)^\gamma$. Regardless of how we parameterize the stellar mass dependence, $\gamma$ is consistently close to three; we thus set $\gamma = 3$. Residuals of the best-fit model correlate with $D_n4000$. Thus, we make the growth rate a function of stellar mass and $D_n4000$. The $D_n4000$ normalization is consistently near 1.5 so we fix it to this value. With the inclusion of $D_n4000$, the dependence on stellar mass is negligible. We remove the explicit stellar mass dependence in the growth rate and arrive at the form in Equation \ref{eq:m_acc_rate}. 

We expect that the growth rate should depend on the stellar mass and environment which appear to be two dominant drivers of galaxy evolution \citep[e.g.,][]{Peng2010}. We suggest the dependence of the growth rate on $D_n4000$ is a consequence of underlying correlations between $D_n4000$, stellar mass and environment \citep{Kauffmann2004}; these correlations likely explain why the growth rate does not need to explicitly depend on these quantities. Later we show that the growth rate does scale with stellar mass. Apparently $D_n4000$ in our model adequately quantifies the dependence of the growth rate on other galaxy properties which may drive the evolution, whatever those properties may be. An ancillary benefit of parameterizing the growth rate as a function of $D_n4000$ is that it mitigates model dependencies and systematic uncertainties in using $D_n4000$ as an evolutionary link.

We parameterize the change in size and stellar velocity dispersion as a power law function of the change in stellar mass. This particular parameterization was motivated by previous work \citep[e.g.,][]{Bezanson2009, Naab2009, Newman2012, Oser2012, Nipoti2012, Hilz2013}. We explored parameterization of $\alpha$ and $\beta$ as a function stellar mass. There does not appear to be any strong dependence of these parameters on stellar mass. However, exploration of such a model may be warranted with larger samples where statistical uncertainties are smaller.

\begin{figure*}
\begin{center}
\includegraphics[width = 2 \columnwidth]{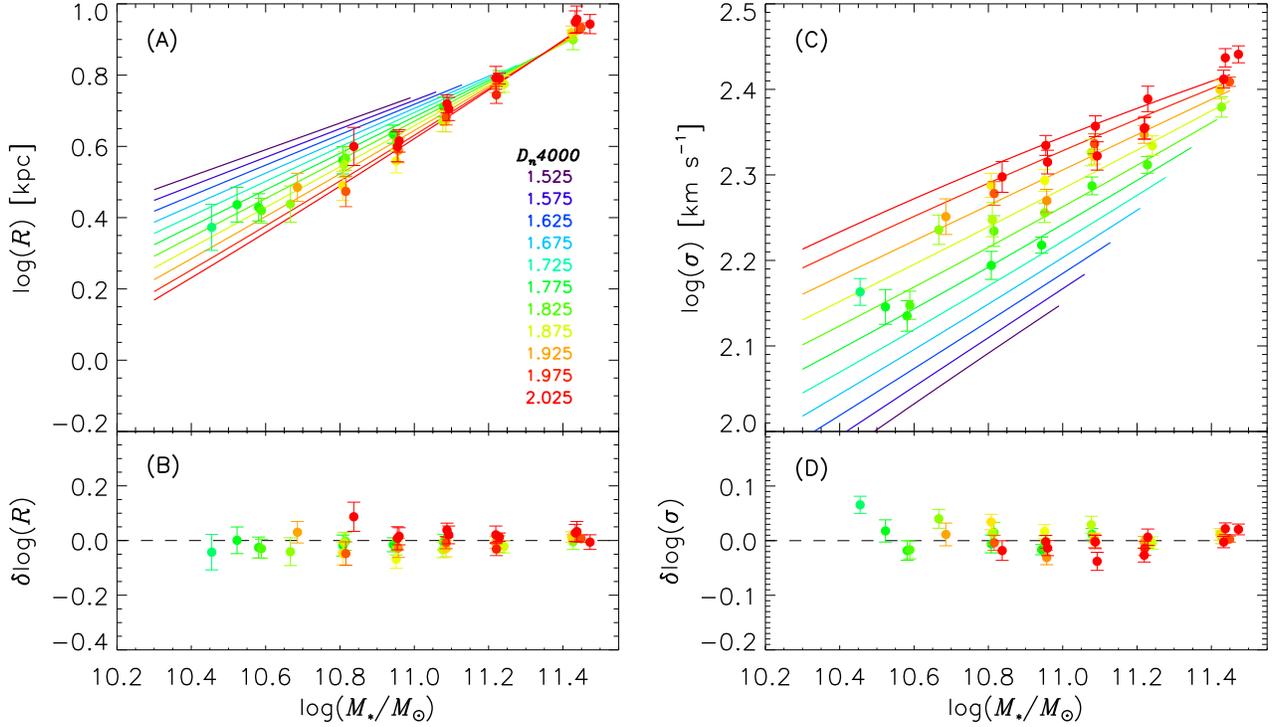}
\end{center}
\caption{Binned properties of SHELS galaxies evolved according to our best-fit model. {Individual SHELS galaxies are evolved according to the procedure described in Section 5.1 and then binned in $M_\ast^e$ and $D_n4000^e$.} (A) Median $R^e$ in bins of $M_\ast^e$ and $D_n4000^e$ of SHELS galaxies (points with error bars) compared to the SDSS sample (colored lines). Colored lines are fits to the SDSS data (Equation \ref{eq:sdss_radius}). The colors of both the lines (SDSS) and the data points (SHELS) reflect the $D_n4000$ value as indicated in the legend, e.g., red lines and points correspond to galaxies with $D_n4000\sim2$. (B) Difference between $R^e$ from SHELS and $R$ from SDSS. (C) Median $\sigma^e$ in bins of $M_\ast^e$ and $D_n4000^e$ of SHELS galaxies (points with error bars) compared to the SDSS sample (colored lines). Colored lines are fits to the SDSS data (Equation \ref{eq:sdss_sigma}). (D) Difference between $\sigma^e$ from SHELS and $\sigma$ from SDSS. Evolved SHELS properties are consistent with SDSS (cf. Figure \ref{fig:vdisp_rad_F2}).}
\label{fig:vdisp_rad_F2_evol}
\end{figure*}

\section{Fitting Our Model and Testing for Consistency}

We describe our fitting procedure and derive our best-fit parameters in Section 5.1 and use the projected SDSS sample to demonstrate consistency of our model fits in Section 5.2.

\subsection{Forward-Fitting Our Model with SHELS and SDSS}

We forward-fit our model using the \citet{Nelder1965} downhill simplex method implemented in \emph{amoeba.pro} in {\sc IDL}. To ensure identification of a global minimum, we perform the fit 1000 times, starting from random initialization of the parameters analogous to the basin hopping method \citep{Wales1997}. The best-fit solution is found within the first few initializations. We derive the confidence intervals for our parameters by bootstrapping the sample. For each bootstrapped sample, we  refit the model using the best-fit parameters as the initial guess and derive the confidence intervals from the distribution of the refitted parameters.

We evolve {each individual} SHELS galaxy to the redshift of the SDSS sample according to the model described in Section 4.4. We bin the evolved SHELS measurements to mitigate the impact of outliers and compare the binned results to SDSS measurements. Figure \ref{fig:vdisp_rad_F2_evol} shows the best-fit results and illustrates the procedure. In detail, we: 
\begin{enumerate}
\item Adopt values for $\epsilon$, $\alpha$ and $\beta$ and determine $\Delta t$ by calculating the time elapsed between each SHELS galaxy and the SDSS sample using the difference in redshift\footnote{$\Delta t$ is the elapsed cosmic time between an individual SHELS galaxy and the median redshift of the SDSS sample at the evolved stellar mass of the SHELS galaxy. The median redshift of the SDSS sample varies with stellar mass (see Figure \ref{fig:sdss_selection}B and Section 2.6.1). $\Delta t$ in step 1 is calculated from the difference between the median redshift of the SDSS sample at a stellar mass $M_\ast^e$ and the SHELS galaxy. However, $M_\ast^e$ depends on $\Delta t$ vis-a-vis Equation \ref{eq:mprime}. Thus, we initially evolve SHELS galaxies to $z=0.1$ but then iterate steps $1-3$ until the change in log($M_\ast^e$) is $<0.01$ dex.}.

\item Calculate $M_\ast^e$, $R^e$ and $\sigma^e$ for each SHELS galaxy using Equations \ref{eq:mprime}, \ref{eq:rprime} and \ref{eq:sprime}, respectively.

\item Calculate $D_n4000^e$ for each SHELS galaxy from $D_n4000$, $\Delta t$ and the $D_n4000$ model (Figure \ref{fig:dn4000_model}).

\item Sort SHELS galaxies into six equally populated bins of $M_\ast^e$ and then sort data in $M_\ast^e$ bins into six equally populated bins of $D_n4000^e$.

\item Calculate the median log($M_\ast^e$)$_{i}$, $D_n4000^e$$_{i}$, log($R^e$)$_{i}$ and log($\sigma^e$)$_{i}$ in each of the 36 bins along with bootstrapped errors $\delta_{R}$$_{i}$ and $\delta_{\sigma}$$_{i}$. Here the subscript $i$ denotes the binned data.

\item Calculate log($R_{SDSS}$)$_{i}$ and log($\sigma_{SDSS}$)$_{i}$ using Equations \ref{eq:sdss_radius} and \ref{eq:sdss_sigma} for the median log($M_\ast^e$)$_{i}$ and $D_n4000^e$$_{i}$.

\item Calculate 
\begin{equation}
\begin{split}
\chi_{R \sigma}^2 =  \sum_{i=1}^{36} & \left[ \frac{ \mathrm{log}(R^e)_i - \mathrm{log}(R_{SDSS})_i  }{\delta_{Ri}} \right]^2 +  \\
& \left[ \frac{ \mathrm{log}(\sigma^e)_i - \mathrm{log}(\sigma_{SDSS})_i  }{\delta_{\sigma i}} \right]^2.
\end{split}
\end{equation}

\item Minimize $\chi_{R \sigma}^2$ to determine best-fit parameters.

\end{enumerate}

We calculate a pseudo reduced-$\chi^2$ by dividing the best-fit $\chi_{R \sigma}^2$ by 66. Here a factor of two comes from summing over two independent observables and a factor of 33 is the number of degrees of freedom (36 data points and 3 fit parameters). The pseudo reduced-$\chi^2$ = 1.48 of the best-fit quantifies the goodness of fit and indicates the model reproduces the data well. The quality of the fit can also be judged from the residuals plotted in Figure \ref{fig:vdisp_rad_F2_evol}. Table \ref{tab:param} gives the best-fit parameters and their confidence intervals.

\subsection{Test of Consistency}

\begin{table}
\begin{center}
\caption{Best-Fit Parameters and Confidence Intervals}
\begin{tabular}{c c c c c c c}
\hline
\hline
Parameter & Best & 2.5\% & 16\% & 50\% & 84\%  & 97.5\% \\
\hline
$\epsilon$ --- Eq. \ref{eq:m_acc_rate} &16.2 & 10.1 & 14.7 & 18.3 & 22.2 & 25.9 \\
$\alpha$ --- Eq. \ref{eq:rprime} & 1.50 & 1.23 & 1.30 & 1.42 & 1.57 & 1.91 \\
$\beta$ --- Eq. \ref{eq:sprime} & 0.41 & 0.31 & 0.34 & 0.37 & 0.42 & 0.50 \\ 
\hline
\label{tab:param}
\end{tabular}
\end{center}
\end{table}

In Section 4.3 we showed that the properties of quiescent galaxies are inconsistent with purely passive evolution. We projected the SDSS population back in time by selecting SDSS samples with larger $D_n4000$ cuts and assuming that $M_\ast$, $R$ and $\sigma$ do not change. As a test of self-consistency, we now project the SDSS population back in time but apply our best-fit model to the SDSS data. We denote all projected SDSS quantities devolved according to our model with the superscript ``p".

\begin{figure*}
\begin{center}
\includegraphics[width = 2 \columnwidth]{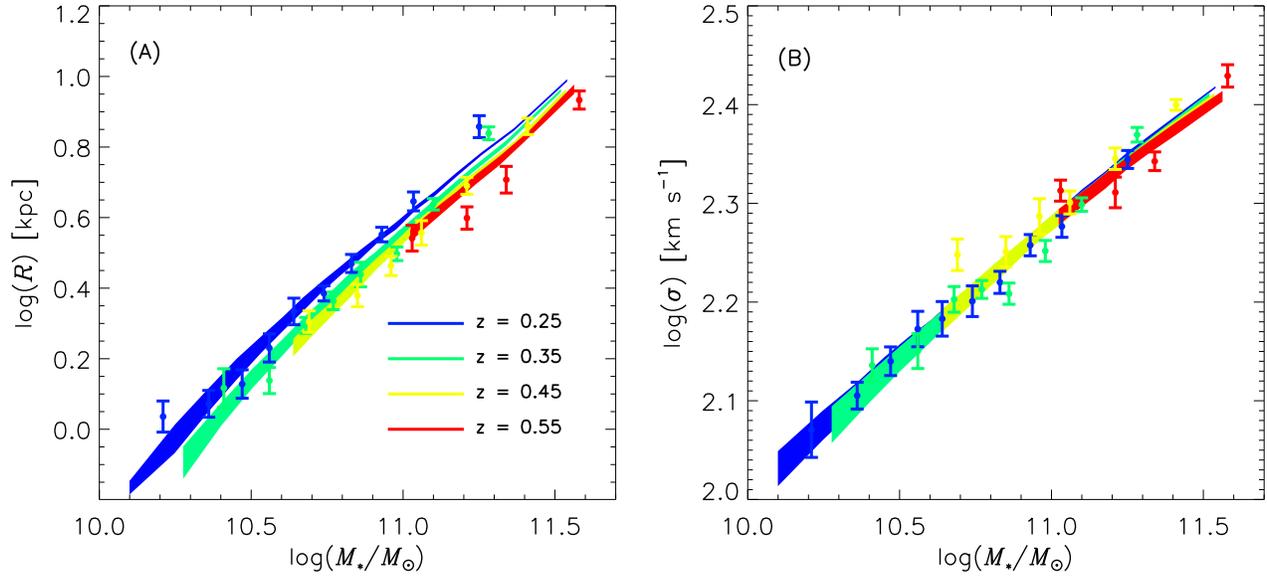}
\end{center}
\caption{(A) $MS$ and (B) $M\sigma$ relations for the SDSS population projected back in time (solid curves) compared to direct measurements from SHELS (points with error bars). The colored shaded regions are the median (A) $R^p$ and (B) $\sigma^p$ in bins of $M_\ast^p$ from SDSS projected to different redshifts. The width of the shaded regions indicate the 95\% confidence interval of the model parameters (Table \ref{tab:param}). The points are the median (A) $R$ and (B) $\sigma$ in bins of $M_\ast$ from SHELS. Error bars are bootstrapped. The projected relations derived after {devolving properties of individual SDSS galaxies according to our best-fit model} are more consistent than ones derived assuming purely passive evolution (cf. Figure \ref{fig:size_vdisp_noevol}). }
\label{fig:size_vdisp_evol}
\end{figure*}

Figure \ref{fig:size_vdisp_evol} shows the $MS$ and $M\sigma$ relations as a function of redshift for the projected SDSS sample compared to direct measurements from SHELS. The two samples are now consistent.

\begin{figure}
\begin{center}
\includegraphics[width =  \columnwidth]{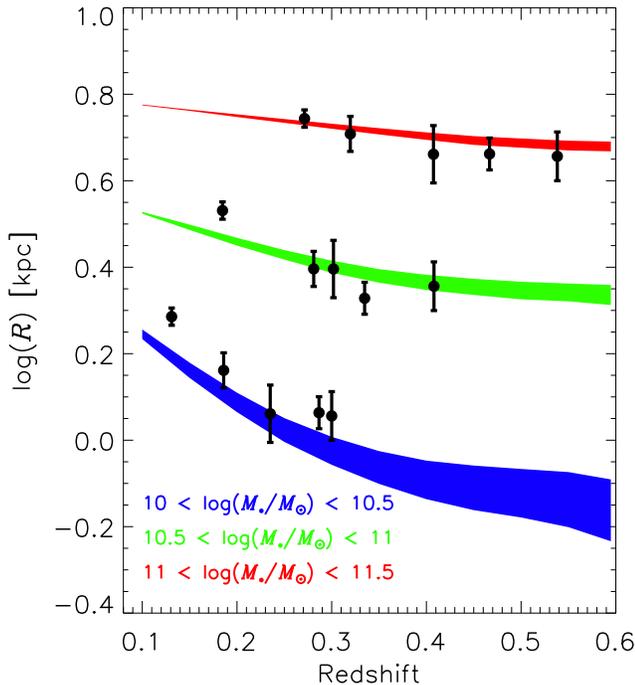}
\end{center}
\caption{Median half-light radii in bins of redshift for galaxies in three stellar mass ranges. The shaded regions are the median $R^p$ calculated in fixed bins of $M_\ast^p$ from SDSS projected to different redshifts. The width of the shaded regions indicate the 95\% confidence interval of the model parameters (Table \ref{tab:param}). The points are $R$ in bins of $M_\ast$ from SHELS. For purposes of this comparison, we include SHELS galaxies down to $z=0.1$. Error bars are bootstrapped.} 
\label{fig:size_evol_mass}
\end{figure}

\citet{Damjanov2019} analyze the half-light radius evolution of quiescent galaxies as a function of redshift and stellar mass. They report differential evolution in galaxy half-light radii with respect to stellar mass; all quiescent galaxies grow but less massive objects grow more. Figure \ref{fig:size_evol_mass} shows the mass dependent evolution of galaxy half-light radii as a function of redshift for the projected SDSS sample and direct measurements from SHELS. Our growth model reproduces the mass dependent evolutionary trends reported by \citet{Damjanov2019}. 

We note that the stellar mass bins in Figure \ref{fig:size_evol_mass} are fixed but individual galaxies grow in time; thus we are not tracing the same objects---in a statistical sense---at various redshifts. Galaxies move from lower mass bins to higher mass bins as they evolve. Thus, our self-consistent modeling is a more nuanced analysis of the growth of quiescent galaxies that connects different mass bins at different epochs. 

Figures \ref{fig:vdisp_rad_F2_evol}, \ref{fig:size_vdisp_evol} and \ref{fig:size_evol_mass} demonstrate that our model of galaxy growth is broadly consistent with observations of massive quiescent galaxies at $z<0.6$.

\section{Model Assumptions}

\begin{figure*}
\begin{center}
\includegraphics[width =  2 \columnwidth]{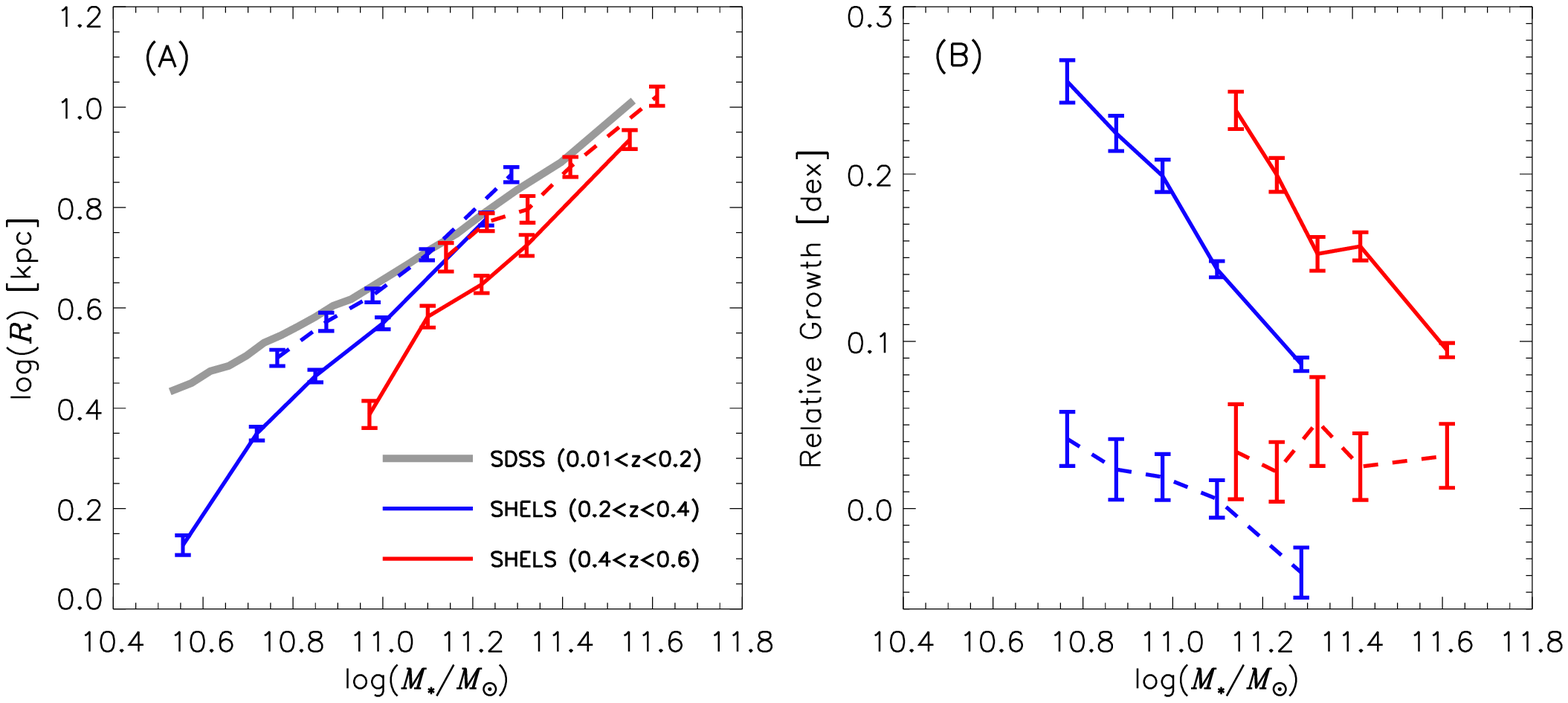}
\end{center}
\caption{(A) $MS$ relation for SDSS (solid gray curve), SHELS at $0.2<z<0.4$ (solid blue curve) and SHELS at $0.4<z<0.6$ (solid red curve). The solid gray curve is the median $R$ in bins $M_\ast$ from SDSS. Solid blue and red curves and points are the median $R$ in bins of $M_\ast$ from SHELS. Dashed curves and points are the median $R^e$ in bins of $M_\ast^e$ from SHELS. The evolved SHELS relations (dashed curves) account for intrinsic growth of individual galaxies but do not include newly quiescent objects appearing at intervening redshifts. (B) Solid curves are the median $\Delta \mathrm{log}(R)$ (see Equation \ref{eq:rprime}) in bins of $M_\ast^e$ from SHELS. These curves quantify intrinsic growth. Colors correspond to samples in (A). The dashed curves are calculated as the difference between the SDSS relation---solid gray curve in (A)---and the evolved SHELS sample---dashed curves in (A). The dashed curves in (B) quantify the impact of progenitor bias. Error bars are bootstrapped.}
\label{fig:pb}
\end{figure*}

We adopt a simple three parameter model to quantify the evolution of quiescent galaxies. Our model assumes a single, mass-independent power law relation between the change in half-light radius and stellar mass and the change in stellar velocity dispersion and stellar mass (Equations \ref{eq:rprime} and \ref{eq:sprime}, respectively). The SHELS sample we analyze contains 3644 galaxies; it is the largest complete sample to its depth with stellar masses, $D_n4000$s, half-light radii and stellar velocity dispersions. Still, the uncertainties of our analysis are dominated by statistical errors and more complex models are unwarranted. As larger data sets become available, more sophisticated models may provide additional insight into the growth of quiescent galaxies.

We use $D_n4000$ as an evolutionary link between galaxies in the local universe and their progenitors at earlier times. $D_n4000$ is a directly measured spectroscopic property which is typically observed at high signal-to-noise ratios and is insensitive to reddening. These features make $D_n4000$ an easy to measure proxy which is straightforward to interpret. 

A fundamental assumption of our model is that $D_n4000$ evolves passively as quantified in Figure \ref{fig:dn4000_model}. Departures from passive evolution caused by in situ star formation and/or mergers could potentially invalidate this assumption. However, the $D_n4000$ in this study is a luminosity weighted quantity measured in a fiber centered on the galaxy and thus is a property of the central stellar population. Star formation at the center of massive quiescent galaxies is rarely observed and the typical accreted mass fraction implied by our model is relatively small ($\lesssim0.2$ dex; see Figure \ref{fig:delta_mass} below) and likely deposited in galaxy outskirts \citep[see also][]{Choi2014}. Thus, deviations from passive evolution are likely negligible for our sample and not a significant source of systematic uncertainty.

Galaxy spectra encode significant information about stellar population age. Our philosophy is to use directly measured quantities where possible. It is beyond the scope of this work to explore more elaborate approaches for deriving stellar population ages based on modeling of galaxy spectra. Advances in stellar population synthesis spectral modeling may provide more robust estimates for stellar population ages and metallicities once systematic uncertainties are better understood \citep{Conroy2013b}. These advances combined with larger spectroscopic samples with high signal-to-noise ratios allowing for measurements of stellar population age and metallicity may justify the use of more sophisticated modeling approaches in the future. However, our simple model already provides important new insights into the coevolution of galaxies and dark matter halos.

\section{Growth of Quiescent Galaxies}

Massive quiescent galaxies at $z<0.6$ do not evolve solely by passive evolution \citep[see Figures \ref{fig:size_vdisp_noevol}, \ref{fig:vdisp_rad_F2} and][]{Zahid2017a}. We quantify their evolution by simultaneously constraining growth of stellar mass, $D_n4000$, half-light radius and stellar velocity dispersion. These joint constraints mitigate degeneracy in our model and provide a robust quantification of quiescent galaxy evolution. Constraints on stellar velocity dispersion evolution are particularly important because they allow us to investigate the coevolution of galaxies and their dark matter halos in Section 8.

Here we interpret our model results to explore the physical mechanisms governing quiescent galaxy evolution. In Section 7.1 we quantify the impact of progenitor bias and in Section 7.2 we show that our model is consistent with minor merger driven growth. We make predictions for how galaxy profiles evolve in Section 7.3 and discuss stellar growth rates in Section 7.4.

\subsection{Impact of Progenitor Bias}

The $MS$ relation of quiescent galaxies evolves as a function of redshift \citep{Shen2003, Trujillo2004a, Zirm2007, Toft2007, Buitrago2008, Guo2009, vanDokkum2010a, vanderwel2014, Lange2015, Roy2018, Damjanov2019}. The origin of this evolution is not well understood partly because both the growth of individual objects and the addition of larger objects at later times (progenitor bias) may contribute to the average growth of quiescent galaxies \citep[e.g.,][]{vanDokkum2001, White2007, Bezanson2009, Naab2009, vanderwel2009, Newman2012, Carollo2013, Damjanov2019}.  A major source of uncertainty is the inability to link galaxies between epochs and identify those which cease star formation in the interim.


We aim to quantify the impact of the continuous addition of quiescent galaxies to the population. Progenitor bias arises because galaxies which become quiescent at later times, tend to be larger and thus contribute to the observed evolution in the $MS$ relation. Our approach is to evolve SHELS galaxies to the redshift of SDSS and compare the $MS$ relation for the two samples. Only galaxies that are already quiescent at $0.2<z<0.6$ are included when we calculate the evolved SHELS $MS$ relation. Thus, any difference between the SDSS $MS$ relation and the evolved SHELS relation is due to new quiescent galaxies which would not be in SHELS.

Figure \ref{fig:pb}A shows the $MS$ relation for SDSS, SHELS and the evolved SHELS samples. Figure \ref{fig:pb}B quantifies the contribution of the intrinsic growth of individual galaxies and progenitor bias on the $MS$ relation. We estimate a small, but statistically significant contribution from progenitor bias. For the stellar mass and redshift range probed by the SHELS sample, we find that the $MS$ relation increases by $0.1-0.25$ dex on average due to growth of individual galaxies; progenitor bias contributes an additional $\sim0.03$ dex. Thus, we conclude that galaxies that become quiescent at later times are indeed on average larger. However, the effect on the average $MS$ relation for the mass and redshift range we probe is small.

Figure \ref{fig:pb} highlights the importance of self-consistently modeling the growth of individual quiescent galaxies. Simply comparing the directly measured $MS$ relations at different redshifts without accounting for growth in both the stellar mass and half-light radius could lead to a biased estimate of progenitor bias.

Stellar mass functions indicate that build-up of the quiescent galaxy population is mass-dependent \citep{Arnouts2007, Pozzetti2010, Ilbert2013, Moustakas2013}; at $z<1$, the number of less massive quiescent galaxies increases rapidly. The evolved SHELS sample probes quiescent galaxies with stellar masses $\gtrsim 10^{10.8} M_\odot$ where the impact of progenitor bias is not large. The effect of progenitor bias appears to be significantly larger for less massive systems \citep{Damjanov2019} and at higher redshifts \citep{Belli2015}. Spectroscopically complete samples to greater depth will provide observations of galaxies over a broader range of stellar masses and at higher redshifts thus enabling a more thorough characterization of the impact of progenitor bias.

\subsection{Minor Merger Driven Growth}

We parameterize the change in half-light radius as $\alpha = \Delta \mathrm{log}(R) / \Delta \mathrm{log} (M_\ast)$. This parameterization has been previously explored in the literature. Theoretically, $\alpha$ depends on the mass ratio of merging systems. For equal mass mergers of spheroids, the change in radius is nearly proportional to the change in mass, i.e. $\alpha \sim 1$ \citep{Hernquist1993}. Theoretical arguments based on the virial theorem and N-body simulations show that for dissipationless (i.e. gas-free or ``dry") minor mergers, size evolution is more efficient, i.e. $\alpha > 1$ \citep{Naab2009, Hopkins2009b, Bezanson2009, Nipoti2009b, Oser2012, Nipoti2012}. For example, \citet{Hopkins2009b} report $\alpha = 1.4-1.8$ and \citet{Nipoti2012} report $\alpha = 1.6 \pm 0.36$. Our best-fit model yields $\alpha = 1.50$; a value consistent with simulations. 

Our results are also consistent with observational estimates based on deep imaging. \citet{Newman2012b} empirically constrain size growth. They estimate the minor merger rate by identifying companions of massive quiescent galaxies from HST Wide Field Camera 3 imaging. Motivated by numerical simulations, they adopt $\alpha = 1.6$ and show that the number of faint companions can account for the growth of quiescent galaxies via minor mergers at $z<1$. 

\begin{figure}
\begin{center}
\includegraphics[width =  \columnwidth]{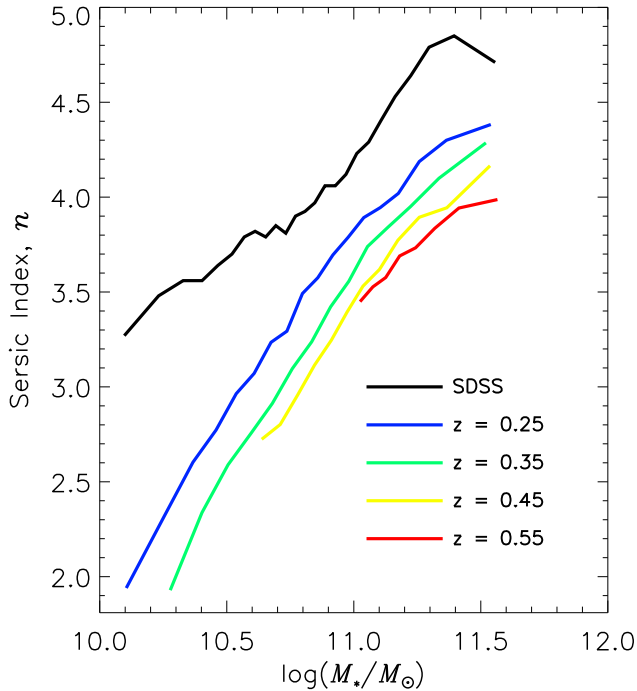}
\end{center}
\caption{Prediction of galaxy profile evolution assuming quiescent galaxies evolve in virial equilibrium and stellar mass is directly proportional to dynamical mass. We project SDSS galaxies back in time to derive the prediction. We calculate $K_d$ from $M_\ast^p$, $R^p$ and $\sigma^p$ and the virial equilibrium relation in Equation \ref{eq:virial}. \citet{Zahid2017a} give $K_d$ as a function of S\'ersic index and we invert this relation to derive the predictions of S\'ersic index evolution.}
\label{fig:sersic}
\end{figure}

\begin{figure*}
\begin{center}
\includegraphics[width =  2\columnwidth]{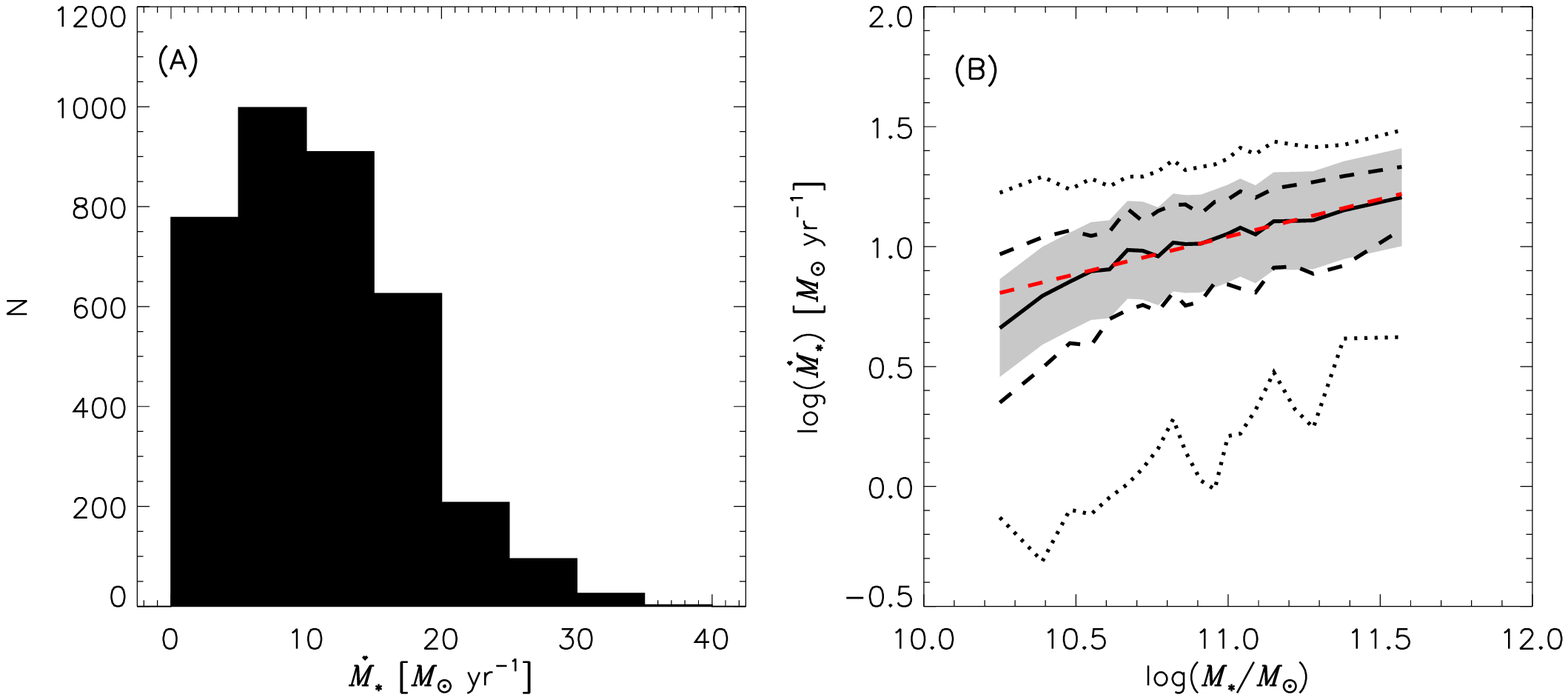}
\end{center}
\caption{(A) Distribution of growth rates, $\dot{M}_\ast$, of SHELS galaxies (see Equation \ref{eq:m_acc_rate}). (B) $\dot{M}_\ast$ as a function of $M_\ast$ for SHELS. The solid black line is the median $\dot{M}_\ast$ in bins of $M_\ast$ and the dashed and dotted black curves indicate limits of the central 50 and 90\% of the galaxies, respectively. The gray band denotes the 95\% confidence interval of our model. The red dashed line is a fit to the median relation (Equation \ref{eq:stellar_rate}).}
\label{fig:accretion_rate}
\end{figure*}

\citet{Newman2012} find that $\sim15\%$ of massive quiescent galaxies in their sample have companions with stellar mass ratios $>0.1$. A preliminary examination of Hyper Suprime-Cam images of SHELS galaxies suggests that a similar fraction of quiescent galaxies show some signature of interaction. A detailed investigation of the imaging data is beyond the scope of this work, but a spectrophotometric correlation analysis over a large area would provide strong constraints for the evolution of quiescent galaxies at $z\lesssim1$.

Our derivation of $\alpha$ is based on a minimal set of assumptions (see Section 6); it is consistent with, but completely independent of previous theoretical and observational studies. Thus, in accordance with previous work, we conclude that minor merger driven growth is the dominant mechanism accounting for the evolution of massive quiescent galaxies at $z<0.6$. 

\subsection{Evolution of Galaxy Profiles}

We parameterize the change in stellar velocity dispersion as $\beta = \Delta \mathrm{log}(\sigma) / \Delta \mathrm{log} (M_\ast)$. Our best-fit model yields $\beta = 0.41$. Stellar velocity dispersion increases as massive quiescent galaxies galaxies grow. There is no theoretical consensus on how stellar velocity dispersion evolves. Several studies conclude that stellar velocity dispersion decreases as galaxies grow via mergers \citep[e.g.,][]{Naab2009, Bezanson2009, Nipoti2012} while others show an increase in stellar velocity dispersion \citep{Hopkins2009b, Hopkins2010}. The reason for this discrepancy is not clear but may be related to changes in galaxy structure.

While quiescent galaxies in the local universe appear to be in virial equilibrium, they are not homologous systems \citep{Trujillo2004a, Zahid2017a}. Thus, conclusions regarding the evolution of stellar velocity dispersion based on the virial theorem which do not explicitly account for evolution in galaxy structure may be oversimplified \citep[e.g,][]{Naab2009, Bezanson2009}. Observations and simulations suggest that galaxy profile shapes evolve as a consequence of minor mergers \citep{vanDokkum2010a, Hilz2013}. Accounting for structural changes may in part explain the discrepant conclusions regarding stellar velocity dispersion evolution.

\begin{figure*}
\begin{center}
\includegraphics[width =  2\columnwidth]{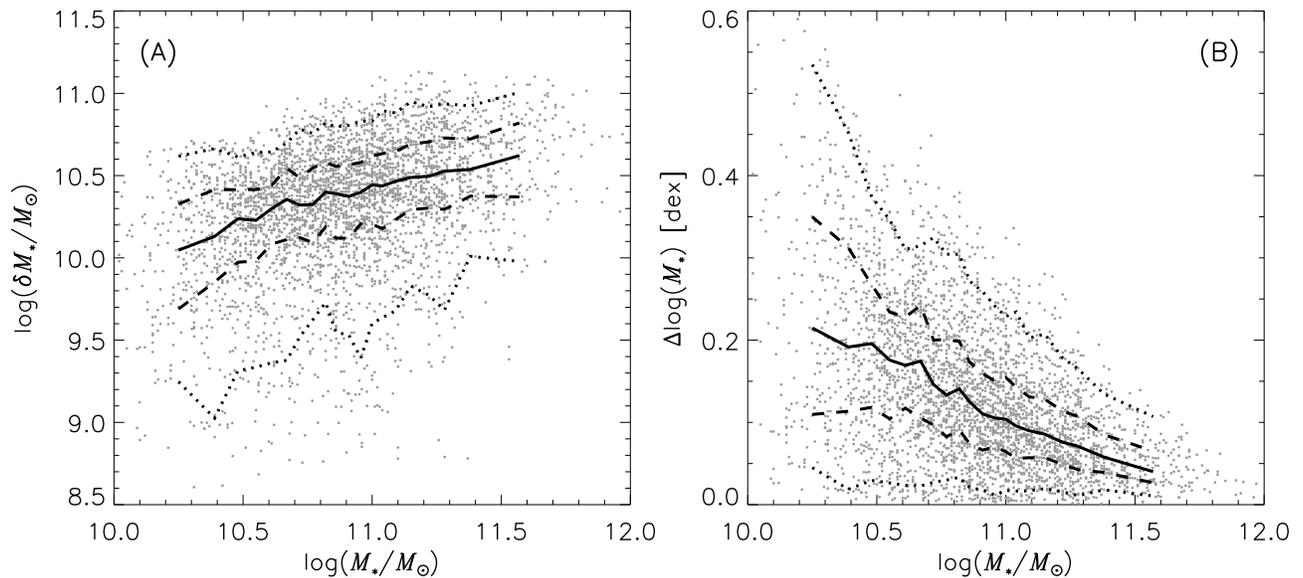}
\end{center}
\caption{(A) Total stellar mass growth, $\delta M_\ast$, as a function of $M_\ast$ for SHELS galaxies (see Equation \ref{eq:delta_m}). (B) Relative stellar mass growth, $\Delta \mathrm{log}(M_\ast)$, as a function of $M_\ast$ for SHELS galaxies. The solid black lines in (A) and (B) are the median $\delta M_\ast$ and $\Delta \mathrm{log}(M_\ast)$ in bins of $M_\ast$, respectively. The dashed and dotted black curves in (A) and (B) indicate limits of the central 50 and 90\% of SHELS galaxies, respectively.}
\label{fig:delta_mass}
\end{figure*}

Assuming quiescent galaxies remain in virial equilibrium as they evolve, our model predicts the evolution of galaxy structure as a function of time. Starting from the scalar virial theorem, we get
\begin{equation}
M_d = K_d \frac{\sigma^2 R}{G}.
\label{eq:virial}
\end{equation}
Here $M_d$ is the dynamical mass, $G$ is the gravitational constant and $\sigma$ and $R$ are the stellar velocity dispersion and half-light radius of the system, respectively \citep{Zahid2017a}. $K_d$ accounts for galaxy structure and projection effects and can be parameterized as a function of the S\'ersic index \citep{Ciotti1997, Prugniel1997, Bertin2002, Zahid2017a}; $K_d$ is anti-correlated with S\'ersic index. 

The dynamical mass is directly proportional to stellar mass for galaxies in SDSS \citep{Taylor2010b, Zahid2017a}. Assuming that a similar proportionality is valid for the SHELS sample, we predict the evolution of galaxy profiles as a function of stellar mass and redshift. For our growth model,
\begin{equation}
\frac{K_d^e}{K_d} = \left( \frac{M_\ast}{M_\ast^e} \right)^{2\beta + \alpha - 1} = \left( \frac{M_\ast}{M_\ast^e} \right)^{\sim1.3}. 
\label{eq:kd}
\end{equation}
$K_d^e$, derived from evolved galaxy properties, decreases as galaxies evolve. Given the anti-correlation between $K_d$ and S\'ersic index, we expect the S\'ersic index to increase as galaxies evolve. This type of evolution is qualitatively consistent with some observations and simulations \citep{vanDokkum2010a, Hopkins2010, Hilz2013, Buitrago2013}.

We can make a detailed prediction for evolution of the S\'ersic index by assuming galaxies evolve in virial equilibrium and stellar mass is directly proportional to dynamical mass. We project the SDSS population back in time as described in Section 4.3 and apply our growth model. This exercise yields stellar mass, half-light radius and stellar velocity dispersion as a function of redshift. We then determine what value of $K_d$ satisfies Equation \ref{eq:virial} and convert it to the S\'ersic index using the relation from \citet{Zahid2017a}. Figure \ref{fig:sersic} shows our prediction.

The current procedure we use to measure the half-light radius of SHELS galaxies does not yield a robust estimate of the S\'ersic index. Thus, Figure \ref{fig:sersic} is a bona fide prediction of our model which can be compared to direct measurements when they become available.

\subsection{The Stellar Growth Rate}

Our model explicitly parameterizes the stellar mass growth rate. Using $\epsilon$ from our best-fit model, we calculate the growth rate for individual SHELS galaxies from Equation \ref{eq:m_acc_rate} using their measured $D_n4000$ and redshift. Figure \ref{fig:accretion_rate}A shows the distribution of growth rates. The median rate of growth is $10  \, M_\odot \, \mathrm{yr}^{-1}$ and $\sim \! 90\%$ of galaxies have growth rates between $1\sim 20 \, M_\odot \, \mathrm{yr}^{-1}$. For a minor merger driven growth interpretation, this rate is a time averaged value quantifying stellar growth via discrete accretion events. Figure \ref{fig:accretion_rate}B shows the growth rate as a function of stellar mass. For massive quiescent galaxies at $0.2 < z < 0.6$, we find a median growth rate of
\begin{equation}
\mathrm{log} \left( \frac{\dot{M}_\ast}{M_\odot \, \, \mathrm{yr}^{-1}} \right) = (1.04 \pm 0.01) + (0.31 \pm 0.02) M_{\ast,11}
\label{eq:stellar_rate}
\end{equation}
where $M_{\ast,11} = \mathrm{log}(M_\ast/10^{11} M_\odot)$. Here statistical uncertainties reflect sample size; model parameters uncertainties are not propagated to the fit.

The growth rate in our model is explicitly a function of redshift and $D_n4000$ (see Equation \ref{eq:m_acc_rate}). The stellar mass scaling in Equation \ref{eq:stellar_rate} is a consequence of the correlation between stellar mass, redshift and $D_n4000$.

We calculate the total change in mass by integrating the growth rate over the time elapsed between the SHELS galaxies and the SDSS sample. Figure \ref{fig:delta_mass}A and \ref{fig:delta_mass}B show the total and relative stellar mass growth, respectively, as a function of redshift. 

We compare our model results with other studies. \citet{vanDokkum2010a} estimate the growth of massive quiescent galaxies ($M_\ast \sim 10^{11.5} M_\odot$) by selecting objects at a fixed number density. They report $\Delta \mathrm{log}(M_\ast) = -0.15 \Delta z$. For the redshift range probed by SHELS, this rate translates to $\sim0.05$ dex increase in stellar mass. \citet{Newman2012} estimate the growth rate for less massive galaxies ($M_\ast > 10^{10.7} M_\odot$) at $z<1$ by identifying companions of massive quiescent host galaxies \citep[see also][]{Lopez-Sanjuan2012}. Adopting a merger timescale of 1 Gyr and a $\Delta t = 4$ Gyr, \citet{Newman2012} estimate a growth of $\sim0.1$ dex for galaxies with $M_\ast \sim 10^{10.7} M_\odot$. Using Equation 4 of \citet{Ferreras2014} and a $\Delta t = 4$ Gyr yields values of $\sim0.14$ dex for galaxies with ($M_\ast \sim 10^{11} M_\odot$). Various estimates based on independent methods yield values of $\Delta \mathrm{log}(M_\ast) \lesssim 0.1$ dex for galaxies with $M_\ast \gtrsim 10^{10.7} \,\, M_\odot$; these values are broadly consistent with our results.

\begin{figure}
\begin{center}
\includegraphics[width =  \columnwidth]{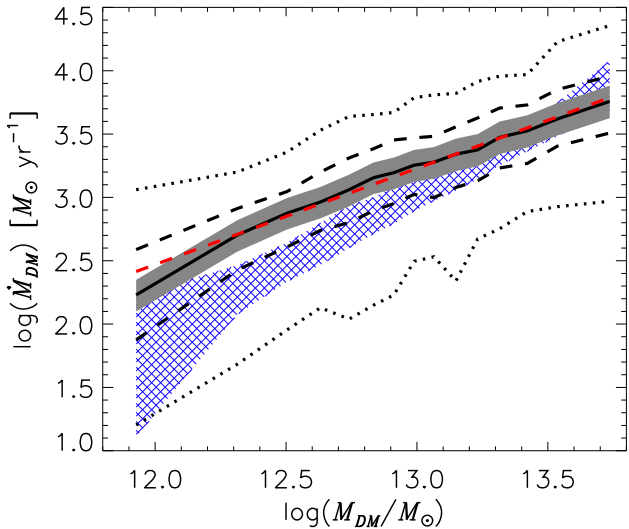}
\end{center}
\caption{Dark matter accretion rate, $\dot{M}_{DM}$, as a function of dark matter halo mass, $M_{DM}$. The solid black line is the median $\dot{M}_{DM}$ in bins of $M_{DM}$ and the dashed and dotted black curves indicate limits of the central 50 and 90\% of SHELS galaxies, respectively. The dark gray band denotes the 95\% confidence interval of our model. The red dashed line is a fit to the median relation (Equation \ref{eq:dmrate}). The blue hashed band is the theoretical $\dot{M}_{DM}$ calculated by \citet{Fakhouri2010} from the Millennium \citep{Springel2005} and Millennium-II \citep{Boylan-Kolchin2009} N-body simulations. The width of the gray band denotes the limits containing 90\% of galaxies. \citet{Fakhouri2010} parameterize $\dot{M}_{DM}$ as a function of $M_{DM}$ and redshift. We calculate an average theoretical $\dot{M}_{DM}$ for each SHELS galaxy (in analogy to the one derived from our model) by averaging the rate calculated for $M_{DM}$ and $M_{DM}^e$. Scatter in the theoretical accretion rate is not reported.}
\label{fig:dark_acc_rate}
\end{figure}

Our results are consistent with minor merger driven evolution of massive quiescent galaxies. In this scenario, the growth rate in our model (Equation \ref{eq:m_acc_rate}) should be interpreted as a time average of discrete events rather than a constant process. For example, we find that the typical growth rate of $10^{11} M_\odot$ galaxies is $\sim \!10 \, M_\odot \, \mathrm{yr}^{-1}$ (see Equation \ref{eq:stellar_rate}). This growth rate could correspond to the accretion of a single $5 \times 10^{10} \, M_\odot$ galaxy or five $10^{10} \, M_\odot$ galaxies in a 5 Gyr timespan. 

\section{The Coevolution of Galaxies and Dark Matter Halos}

\begin{figure*}
\begin{center}
\includegraphics[width = 2 \columnwidth]{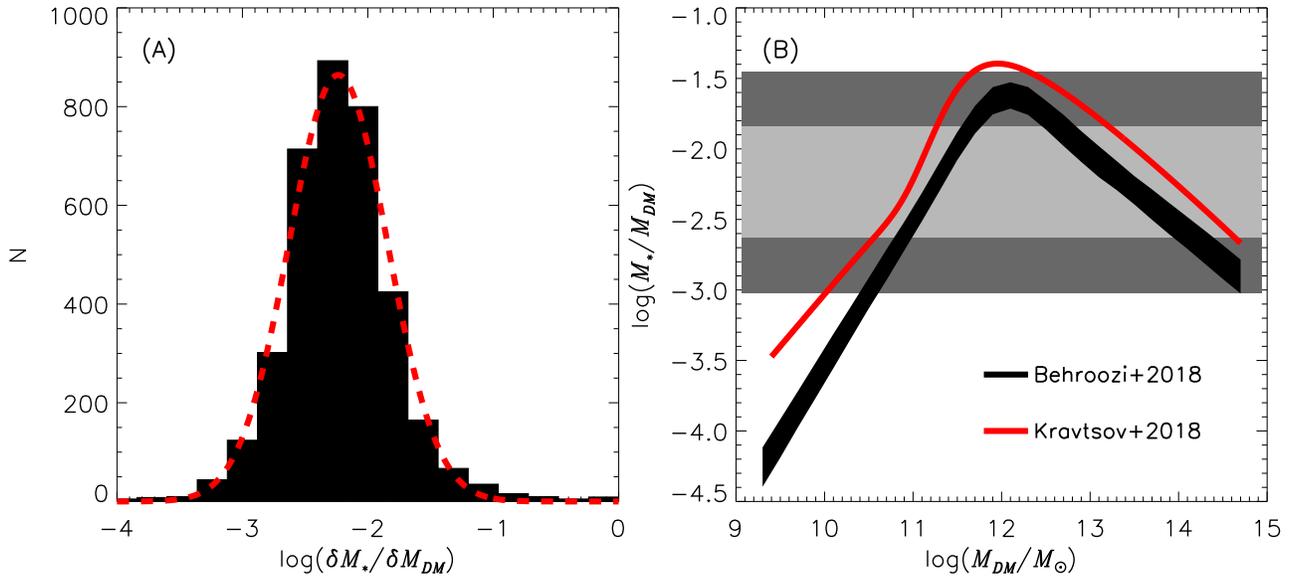}
\end{center}
\caption{(A) Distribution of stellar-to-dark matter ratio for accreted objects, log($\delta {M}_\ast/\delta {M}_{DM})$, based on our model. The red curve is the best-fit Gaussian with a mean and standard deviation of $-2.23 \pm 0.01$ and $0.39\pm0.01$, respectively. The errors are Poisson uncertainties and do not account for uncertainties in our growth model. (B) Stellar mass-halo mass relation from \citet[black curve]{Behroozi2018} and \citet[red curve]{Kravtsov2018}. These relations are derived using independent methodologies. The width of the black curve accounts for the systematic and statistical uncertainties. The gray bands are the 1 and 2$\sigma$ bands corresponding to the distribution in (A). The log($\dot{M}_\ast/\dot{M}_{DM})$ in (A) correspond to stellar-to-dark matter ratios of normal galaxies.}
\label{fig:acc_ratio}
\end{figure*}

Dark matter halos are in virial equilibrium and thus halo velocity dispersion is strongly correlated with halo mass \citep{Evrard2008}. \citet{Zahid2018} analyze the Illustris hydrodynamical cosmological simulation \citep{Vogelsberger2014a} and show that the stellar velocity dispersion is nearly equal to the dark matter halo velocity dispersion confirming previous suggestions \citep{Schechter2015, Zahid2016c}. They derive a relation from the simulations that may be used to infer dark matter halo mass from the observed stellar velocity dispersion. The exact origin of this powerful relation is not fully understood but is likely related to the very nearly isothermal total density profiles of quiescent galaxies \citep[e.g.,][]{Koopmans2009, Remus2013}.

In Section 8.1 we calculate the growth rate of dark matter halos. We use these calculations in Section 8.2 to investigate the coevolution of galaxies and their dark matter halos and to understand the scaling of the stellar mass-halo mass relation based on simulations.

\subsection{Growth Rate of Dark Matter Halos}

The theoretical relation between stellar velocity dispersion and dark matter halo mass from \citet{Zahid2018} is
\begin{equation}
\mathrm{log}\left(  \frac{M_{DM}}{10^{12} ~ M_\odot}   \right) = \eta+ \zeta ~ \mathrm{log}\left(  \frac{\sigma_{sim}}{100 ~\mathrm{km ~ s}^{-1}}   \right). 
\label{eq:sigma_dm}
\end{equation}
Here $\eta =  0.16\pm 0.03$, $\zeta= 3.31 \pm 0.10$ and the relation has 0.17 dex intrinsic scatter. $M_{DM}$ is the dark matter halo mass and $\sigma_{sim}$ is the stellar velocity dispersion which is computed in the simulation in direct analogy to the measurements \citep[for details see][]{Zahid2018}. Thus, Equation \ref{eq:sigma_dm} is compatible with the stellar velocity dispersions we analyze and can be directly used with the observations to infer dark matter halo mass. By combining this relation with our growth model, we investigate the coevolution of galaxies and their dark matter halos.

We emphasize that the theoretical relation in Equation \ref{eq:sigma_dm} results from the nearly direct proportionality between stellar velocity dispersion and dark matter halo velocity dispersion. The simulations merely validate the assertion of this direct proportionality. Simply equating dark matter halo velocity dispersion to stellar velocity dispersion and deriving dark matter halo mass from the relation of virial equilibrium would not significantly impact the analysis that follows. Our empirical analysis of the growth of dark matter halos is essentially independent of the simulations and thus provides a powerful test of the theoretically predicted evolution of dark matter halos.

We compute the average dark matter growth rate of SHELS galaxies as
\begin{equation}
\delta M_{DM} = (M_{DM}^e - M_{DM})
\end{equation} 
and 
\begin{equation}
\dot{M}_{DM} = \delta M_{DM}/\Delta t.
\end{equation} 
Here $M_{DM}$ and $M_{DM}^e$ are dark matter halo masses calculated from $\sigma$ and $\sigma^e$ of SHELS galaxies, respectively, using Equation \ref{eq:sigma_dm}. $\delta M_{DM}$ is the change in dark matter halo mass and $\Delta t$ is the cosmic time elapsed between SHELS galaxies and the SDSS sample. We interpret $\dot{M}_{DM}$ as the average dark matter growth rate for galaxies evolving over the redshift range probed by the SHELS and SDSS samples. As with the stellar growth rate we derive, $\dot{M}_{DM}$ is a time averaged quantity characterizing discrete accretion events.

Figure \ref{fig:dark_acc_rate} shows the dark matter growth rate as a function of dark matter halo mass. The best-fit is
\begin{equation}
\mathrm{log} \left( \frac{\dot{M}_{DM}}{M_\odot \, \, \mathrm{yr}^{-1}} \right) = (2.48 \pm 0.02) + (0.75 \pm 0.02) M_{12},
\label{eq:dmrate}
\end{equation}
where $M_{12} = \mathrm{log}(M_{DM}/10^{12} M_\odot)$. Here statistical uncertainties reflect sample size; model parameters uncertainties are not propagated to the fit.

We compare our empirically derived dark matter growth rate to the average growth rate from N-body simulations \citep[Equation 2 in][]{Fakhouri2010}. We emphasize that our empirical estimate is independent of the N-body simulations. Thus, the consistency in Figure \ref{fig:dark_acc_rate} is astonishing and demonstrates that stellar velocity dispersion is a powerful observable proxy linking galaxies to dark matter halos. We note that the small systematic differences may be due to our quiescent galaxy selection, systematics in the simulations or other unknown issues.

\subsection{The Stellar Mass-Halo Mass Relation}

We calculate the stellar-to-dark matter growth ratio from the stellar and dark matter halo growth rates (see Figures \ref{fig:accretion_rate}B and \ref{fig:dark_acc_rate}, respectively). We interpret this growth ratio as the stellar-to-dark matter ratio of accreted objects. Figure \ref{fig:acc_ratio}A shows that the stellar-to-dark matter growth ratio is log-normally distributed for SHELS galaxies. The typical stellar-to-dark matter growth ratio is log$(\delta {M}_\ast/\delta {M}_{DM}) \sim -2.3$. We compare the distribution of growth ratios to two recent stellar mass-halo mass (SMHM) relations derived from abundance matching \citep{Kravtsov2018} and semi-empirical modeling \citep{Behroozi2018}. Figure \ref{fig:acc_ratio}B shows the stellar-to-dark matter growth ratios are consistent with accretion of objects with $M_{DM} > 10^{10} M_\odot$. 

Figure \ref{fig:acc_ratio}B shows the range of dark matter halos hosting galaxies which satisfy the empirical constraints we derive from the distribution of stellar-to-dark matter growth ratios. However, the merger of two galaxies above the knee of the SMHM relation ($M_{DM} \gtrsim 10^{12} M_\odot$) moves galaxies off the relation. The remnant of such a merger would typically have a stellar-to-dark matter ratio which is larger than the massive progenitor. On the other hand, massive galaxies merging with galaxies below the knee could yield a merger remnant with a stellar-to-dark matter ratio which is smaller than the massive progenitor, thus moving massive galaxies along the relation. 

Figure \ref{fig:acc_ratio} indicates that massive galaxies merge with objects residing below the knee of the SMHM. For example, the typical stellar-to-dark matter growth ratio of log$(\delta M_\ast /\delta M_{DM})\sim-2$ (see Figure \ref{fig:acc_ratio}) corresponds to a galaxy which has $M_\ast \sim10^{9} M_\odot$ and $M_{DM} \sim 10^{11} M_\odot$. Thus, massive quiescent galaxies at $z<0.6$ appear to grow from minor mergers with dark matter halos of mass $10^{10} \lesssim M_{DM} \lesssim 10^{12} M_\odot$. This conclusion may change if diffuse dark matter accretion, i.e. accretion of low mass halos not hosting stars, is substantial.

We calculate the trajectory of galaxies in the SMHM plane. We parameterize the change in stellar velocity dispersion with respect to stellar mass as $\Delta \mathrm{log}(\sigma) \propto \beta \Delta \mathrm{log}(M_{\ast})$ (Equation \ref{eq:sprime}). We derive the change in dark matter halo mass with respect to stellar velocity dispersion from Equation \ref{eq:sigma_dm}: $\Delta \mathrm{log}(M_{DM}) \propto \zeta \Delta \mathrm{log}(\sigma)$. Thus, galaxies evolve such that $\Delta \mathrm{log}(M_{\ast})/\Delta \mathrm{log}(M_{DM}) \propto \frac{1}{\beta \zeta}$. 

\begin{figure}
\begin{center}
\includegraphics[width =  \columnwidth]{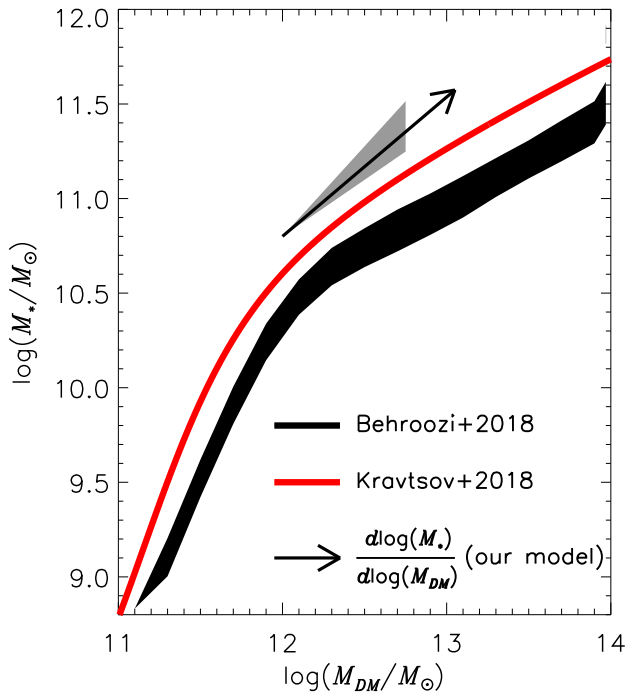}
\end{center}
\caption{$M_\ast$ as a function of $M_{DM}$. The SMHM relations are from \citet[black curve]{Behroozi2018} and \citet[red curve]{Kravtsov2018}. The arrow shows how galaxies evolve according to our model in the SMHM plane. The gray band denotes the 95\% confidence interval accounting for the statistical uncertainties of our model (see Table \ref{tab:param}). Galaxies evolve along the SMHM relation. }
\label{fig:mstar_mhalo}
\end{figure}

Figure \ref{fig:mstar_mhalo} compares our model evolution to the \citet{Kravtsov2018} and \citet{Behroozi2018} SMHM relations. Galaxies and dark matter halos coevolve along the SMHM relation.

The SMHM relation connects galaxies to their dark matter halos. In its simplest form, the relation is constructed by assuming a one-to-one correspondence between the stellar mass derived from observations and halo mass calculated from N-body simulations \citep[for a review see][]{Wechsler2018}. The approach is limited in its ability to elucidate the underlying physics governing galaxies and their dark matter halos \citep[e.g.,][]{Silk2012}. The origin of the scaling is generally interpreted as reflecting the efficiency of star formation which is purportedly set by AGN feedback in massive galaxies \citep{Silk1998, Benson2003, Croton2006}. 

Remarkably, our model of minor merger driven growth reproduces the scaling of the SMHM relation for massive galaxies. Baryonic physical processes like AGN feedback may be necessary to suppress star formation in massive galaxies. However, these feedback mechanisms are not directly responsible for the scaling of the SMHM relation. In the absence of star formation, the turnover and slope of the SMHM relation of massive galaxies appears to result from dry minor merging.

\section{Conclusion}

We empirically model the coevolution of quiescent galaxies and their dark matter halos at $z<0.6$ independent of the simulations. Our three parameter model links quiescent galaxies observed in SDSS and SHELS---two spectroscopically complete, high-quality data sets---by self-consistently quantifying the evolution of stellar mass, $D_n4000$, half-light radius and stellar velocity dispersion. We use stellar velocity dispersion as a proxy of dark matter halo mass and empirically constrain their growth. As galaxies evolve, they move along the stellar mass-halo mass relation based on simulations. We conclude that the coevolution of massive quiescent galaxies and their dark matter halos over the last $\sim \! 6$ billions years results from dry minor merging. Our model clarifies aspects of the stellar mass-halo mass relation based on simulations and is a benchmark for guiding theoretical investigations of the coevolution of galaxies and dark matter halos.

The typical rate of stellar mass growth for our best fit model is $\sim \! 10 \,\, M_\odot \,\, \mathrm{yr}^{-1}$. The best-fit power law index $\alpha = 1.50$ relates the change in half-light radius to the change in stellar mass. Our $\alpha$ is consistent with previous constraints from observations and simulations of galaxy growth by dry minor mergers.

We constrain the evolution of stellar velocity dispersion and derive the best-fit power law index $\beta = 0.41$ which relates the change in stellar velocity dispersion to the change in stellar mass. Our results agree with some simulations though not all theoretical studies yield qualitatively consistent results. Our analysis of stellar velocity dispersion provides an additional dimension for constraining simulations of galaxy evolution and a directly observable link to dark matter halos. 

\citet{Zahid2018} show that stellar velocity dispersion is a direct proxy of dark matter halo mass. We empirically model how stellar velocity dispersion changes as galaxies grow and use these results to constrain the coevolution of galaxies and their dark matter halos. Our analysis is independent of the simulations and thus provides an important test of our theoretical understanding of the evolution of dark matter over the last six billion years.



We derive dark matter growth rates for individual galaxies from our model that are consistent with N-body simulations. Massive quiescent galaxies at $z<0.6$ grow by accreting halos of mass $10^{10} \lesssim M_{DM} \lesssim 10^{12} M_\odot$. Galaxies and dark matter halos coevolve parallel to the relation between stellar mass and halo mass. The turnover and slope of this relation appear to be largely a consequence of hierarchical formation rather than baryonic feedback process which are secondary. 

Our approach is a new route to a deeper understanding of galaxies and their dark matter halos. Our analysis exploits the synergy between data and simulations underscoring the power of complete, homogeneously observed spectrophotometric samples---even ones of modest size. Connecting our model results for quiescent galaxies to the star-forming population will provide a coherent picture of galaxy evolution at intermediate redshifts. Observations of star formation rates at multiple epochs constrain the growth of star-forming galaxies \citep[e.g.,][]{Leitner2012}. Metallicity evolution can also be traced by such models \citep{Zahid2012b, Zahid2013c, Zahid2014b, Zahid2017b}. Stellar metallicity measurements provide a missing link to connect star-forming and quiescent galaxies \citep[e.g.,][]{Peng2015}. Complete spectroscopic samples observed at higher signal-to-noise ratios are necessary to robustly measure stellar metallicities of individual galaxies and forge this connection. The next generation of large multi-object spectrographs on 8m telescopes \citep[e.g., VLT MOONS, Subaru PFS;][]{Cirasuolo2012, Sugai2015} will soon make this possible.



\acknowledgements
 
We appreciate the thoughtful comments of the anonymous reviewer which helped clarify the manuscript. HJZ acknowledge the generous support of the Clay Fellowship. MJG and JS are supported by the Smithsonian Institution. ID acknowledges support from the Canada Research Chair Program. We are grateful to Rolf Kudritzki and Drew Newman for carefully reading the manuscript and providing thoughtful comments. We thank Dan Fabricant and Yousuke Utsumi for measuring velocity dispersions and sizes, respectively, Brandon Lazo for examining the imaging data to identify interacting galaxies and Peter Behroozi for assistance accessing his results. We had stimulating discussions with Andi Burkert, Benedikt Diemer and Charlie Conroy. This work benefitted from the efforts of all of the Subaru Telescope staff and the HSC builders. We thank Dr. Okabe for sharing computer resources necessary for reduction and analysis of HSC images. The creation and distribution of the NYU-VAGC is funded by the New York University Department of Physics. 

Funding for SDSS-III has been provided by the Alfred P. Sloan Foundation, the Participating Institutions, the National Science Foundation, and the U.S. Department of Energy Office of Science. The SDSS-III web site is http://www.sdss3.org/. SDSS-III is managed by the Astrophysical Research Consortium for the Participating Institutions of the SDSS-III Collaboration including the University of Arizona, the Brazilian Participation Group, Brookhaven National Laboratory, University of Cambridge, Carnegie Mellon University, University of Florida, the French Participation Group, the German Participation Group, Harvard University, the Instituto de Astrofisica de Canarias, the Michigan State/Notre Dame/JINA Participation Group, Johns Hopkins University, Lawrence Berkeley National Laboratory, Max Planck Institute for Astrophysics, Max Planck Institute for Extraterrestrial Physics, New Mexico State University, New York University, Ohio State University, Pennsylvania State University, University of Portsmouth, Princeton University, the Spanish Participation Group, University of Tokyo, University of Utah, Vanderbilt University, University of Virginia, University of Washington, and Yale University.

\emph{Facilities:} MMT at Fred Lawrence Whipple Observatory and Subaru Telescope operated by National Astronomical Observatory of Japan.

\section*{Appendix}

\begin{table}
\begin{center}
\caption{Best-Fit Parameters and Confidence Intervals}
\begin{tabular}{c c c c c c c}
\hline
\hline
Parameter & Best & 2.5\% & 16\% & 50\% & 84\%  & 97.5\% \\
\hline
\hline
&\multicolumn{5}{c}{$MLR = 0.30$ dex (N = 3644)}  \\
\hline
$\epsilon$ &16.2 & 10.1 & 14.7 & 18.3 & 22.2 & 25.9 \\
$\alpha$ & 1.50 & 1.23 & 1.30 & 1.42 & 1.57 & 1.91 \\
$\beta$ & 0.41 & 0.31 & 0.34 & 0.37 & 0.42 & 0.50 \\
\hline
&\multicolumn{5}{c}{$MLR = 0.45$ dex (N = 3387)}  \\
\hline
$\epsilon$&18.3 & 11.2 & 15.0 & 19.4 & 23.6 & 30.5 \\
$\alpha$  & 1.48 & 1.20 & 1.30 & 1.42 & 1.58 & 1.85 \\
$\beta$  & 0.39 & 0.31 & 0.35 & 0.39 & 0.44 & 0.51 \\
\hline
&\multicolumn{5}{c}{$MLR = 0.60$ dex (N = 2896)}  \\
\hline
$\epsilon$&10.9 & 5.54 & 7.26 & 9.94 & 12.9 & 18.7 \\
$\alpha$  & 1.87 & 1.46 & 1.71 & 1.95 & 2.31 & 2.83 \\
$\beta$  & 0.54 & 0.41 & 0.49 & 0.58 & 0.70 & 0.87 \\
\hline
&\multicolumn{5}{c}{$MLR = 0.75$ dex (N = 2219)}\\
\hline
$\epsilon$&10.3 & 5.28 & 6.71 & 8.98 & 12.3 & 18.0 \\
$\alpha$  & 1.78 & 1.40 & 1.67 & 1.94 & 2.30 & 2.72 \\
$\beta$  & 0.63 & 0.43 & 0.56 & 0.67 & 0.82 & 1.00 \\
\hline
&\multicolumn{5}{c}{Twice-Solar (N = 3644)}  \\
\hline
$\epsilon$ &14.7 & 6.78 & 9.78 & 14.0 & 19.4 & 25.5 \\
$\alpha$  & 1.50 & 1.18 & 1.32 & 1.53 & 1.81 & 2.17 \\
$\beta$  & 0.50 & 0.39 & 0.47 & 0.55 & 0.68 & 0.86 \\
\hline
\label{tab:param_appendix}
\end{tabular}
\end{center}
\vspace{-5mm}
\begin{tablenotes}
 \small
     \item  {Best-fit parameters and corresponding confidence intervals of the evolutionary model fits. The first set are the fiducial fit described in Section 5.1. The next three sets of parameters are derived varying the $MLR$ in selecting the SHELS sample. The final set of parameters is based on our fiducial SHELS sample selection criteria but $D_n4000$ evolution generated from a twice-solar model input to FSPS as described in Section 4.2. The sample size is provided in parenthesis above each set of parameters.}
    \end{tablenotes}
\end{table}

\begin{figure*}
\begin{center}
\includegraphics[width = 2 \columnwidth]{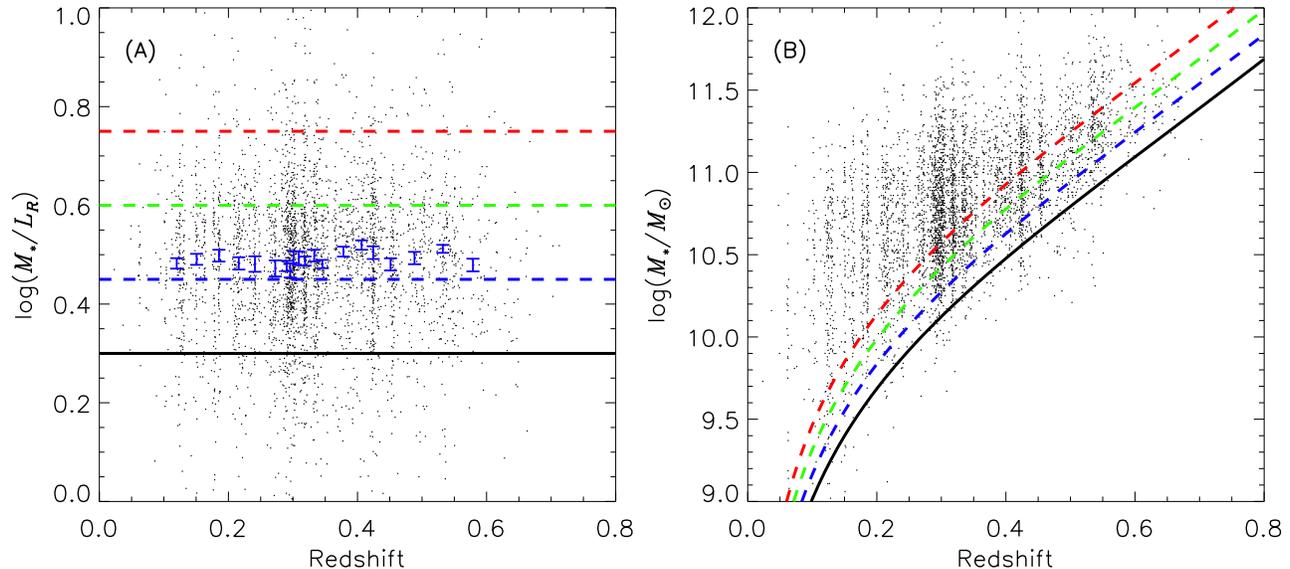}
\end{center}
\caption{ {(A) $R$-band $MLR$ as a function of redshift for the parent SHELS sample of quiescent galaxies. Bootstrapped error bars show the median $MLR$ in 20 equally populated bins of redshift. Over the redshift range we examine, the $MLR$ is independent of redshift. The solid black line is the fiducial $MLR = 0.3$ dex we adopt to translate the SHELS magnitude limit to a stellar mass limit; the blue, green and red dashed lines correspond $MLR$s of 0.45, 0.60 and 0.75 dex, respectively. (B) $M_\ast$ as a function of redshift for the parent SHELS sample of quiescent galaxies. The curves are the stellar mass limit assuming the $MLR$s shown in (A).} }
\label{fig:shels_selection_appendix}
\end{figure*}

\begin{figure}
\begin{center}
\includegraphics[width =  \columnwidth]{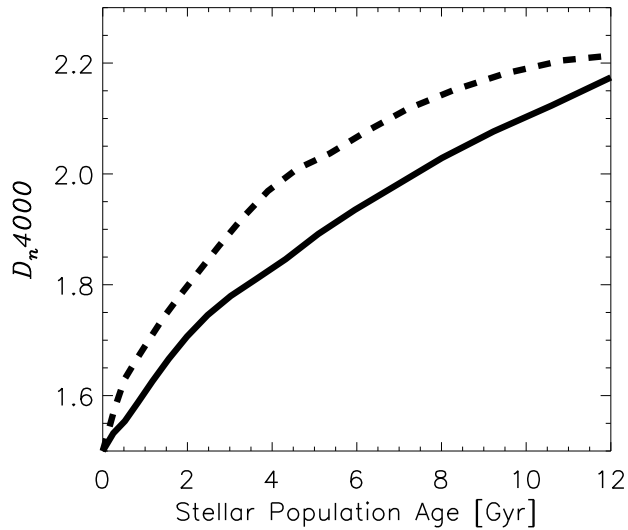}
\end{center}
\caption{ {Evolution of $D_n4000$ as a function of time calculated for the fiducial FSPS model with solar metallicity and constant SFR for 1 Gyr (solid black curve as in Figure \ref{fig:dn4000_model}) and a model with twice-solar metallicity and constant SFR for 1 Gyr (dashed curve).}}
\label{fig:dn4000_model_appendix}
\end{figure}

{We investigate the impact of systematic effects resulting from the selection of the SHELS sample and the FSPS input metallicity adopted to generate our fiducial model of $D_n4000$ evolution. We vary the SHELS sample selection criteria and FSPS input metallicity and fit our evolutionary model (Section 4.4) following the procedure described in Section 5.1. }

{We adopt a $MLR = 0.3$ dex to convert the magnitude limit of the SHELS sample to an approximate stellar mass limit (Section 2.6.2). We test the impact of choosing this $MLR$ by adopting a range of values between 0.3 - 0.75 dex. Figure \ref{fig:shels_selection_appendix}A shows the $MLR$ distribution of SHELS galaxies. The different colored lines indicate various $MLR$s used to approximate the stellar mass limit. Figure \ref{fig:shels_selection_appendix}B shows the resulting stellar mass limits. }

{We apply the various stellar mass limits shown in Figure \ref{fig:shels_selection_appendix}B to select different SHELS samples. We derive the best-fit parameters of our evolutionary model using these samples. The best-fit parameters are in Table \ref{tab:param_appendix}. Results from fitting the various samples are consistent within the 95\% confidence interval. Thus, we conclude that at the level of the statistical uncertainties, our evolutionary model results are insensitive to the $MLR$ used to select the SHELS sample.}

{We connect galaxies observed in SHELS to their descendants in SDSS using the $D_n4000$ index as a stellar population age indicator. In Section 4.2 we derive the theoretical evolution of $D_n4000$ using stellar population synthesis models. We test whether our results are sensitive to the particular choice of FSPS model metallicity. Figure \ref{fig:dn4000_model_appendix} shows the $D_n4000$ index evolution for our fiducial model and for a model with twice-solar metallicity. We derive the best-fit parameters of our evolutionary model using to the twice-solar metallicity model for $D_n4000$ time evolution. The best-fit parameters are provided in Table \ref{tab:param_appendix}. All parameters are consistent within the 95\% confidence interval.}

{We test the impact of systematic effects resulting from selection of the SHELS sample and the FSPS input metallicity adopted to generate our fiducial model of $D_n4000$ evolution. We demonstrate that the best-fit parameters are consistent within the statistical uncertainties. Thus, we conclude that statistical errors and not systematic effects are the dominant source of uncertainty.} 

\bibliographystyle{aasjournal}
\bibliography{/Users/jabran/Documents/latex/metallicity}

 \end{document}